\documentclass[aps,physrev,twocolumn,superscriptaddress,floatfix]{revtex4-2}

\usepackage{graphicx}
\usepackage{dcolumn}
\usepackage{bm}
\usepackage{amsmath,amssymb}

\usepackage{afterpage}
\usepackage[uncertainty-mode = separate]{siunitx}
\let\oldqty\qty
\renewcommand{\qty}[2]{\oldqty{#1}{#2}}

\usepackage[version=4]{mhchem}
\usepackage{ulem}
\usepackage{color}

\usepackage{placeins}
\usepackage[final]{changes}
\definechangesauthor[name = {Reviewer 1}, color=blue]{R1}
\definechangesauthor[name = {Reviewer 2}, color=blue]{R2}

\begin{document}


\title{\replaced[id=R1]{Photorefractive-based on-chip optical power limiter against light-injection attacks in quantum key distribution}{Photorefractive-based optical fuse on chip against light injection attacks on quantum key distribution}}

\author{Min Chen}
\altaffiliation{Contributed equally to this work}
\affiliation{Laboratory of Quantum Information, University of Science and Technology of China, Hefei 230026, China}
\affiliation{Anhui Province Key Laboratory of Quantum Network, University of Science and Technology of China, Hefei 230026, China}

\author{Hong-Yan Song}
\altaffiliation{Contributed equally to this work}
\affiliation{Anhui Asky Quantum Technology CO., LTD, Wuhu 241000, China}

\author{Jia-Lin Chen}
\affiliation{Laboratory of Quantum Information, University of Science and Technology of China, Hefei 230026, China}
\affiliation{Anhui Province Key Laboratory of Quantum Network, University of Science and Technology of China, Hefei 230026, China}

\author{Peng Ye}
\affiliation{Laboratory of Quantum Information, University of Science and Technology of China, Hefei 230026, China}
\affiliation{Anhui Province Key Laboratory of Quantum Network, University of Science and Technology of China, Hefei 230026, China}

\author{Guo-Wei Zhang}
\affiliation{Laboratory of Quantum Information, University of Science and Technology of China, Hefei 230026, China}
\affiliation{Anhui Province Key Laboratory of Quantum Network, University of Science and Technology of China, Hefei 230026, China}

\author{Fang-Xiang Wang}
\affiliation{Laboratory of Quantum Information, University of Science and Technology of China, Hefei 230026, China}
\affiliation{Anhui Province Key Laboratory of Quantum Network, University of Science and Technology of China, Hefei 230026, China}

\author{Li Zhang}
\affiliation{Anhui Asky Quantum Technology CO., LTD, Wuhu 241000, China}

\author{Shuang Wang}
\affiliation{Laboratory of Quantum Information, University of Science and Technology of China, Hefei 230026, China}
\affiliation{Anhui Province Key Laboratory of Quantum Network, University of Science and Technology of China, Hefei 230026, China}
\affiliation{Hefei National Laboratory, University of Science and Technology of China, Hefei 230088, China}

\author{De-Yong He}
\affiliation{Laboratory of Quantum Information, University of Science and Technology of China, Hefei 230026, China}
\affiliation{Anhui Province Key Laboratory of Quantum Network, University of Science and Technology of China, Hefei 230026, China}
\affiliation{Hefei National Laboratory, University of Science and Technology of China, Hefei 230088, China}

\author{Zhen-qiang Yin}
\affiliation{Laboratory of Quantum Information, University of Science and Technology of China, Hefei 230026, China}
\affiliation{Anhui Province Key Laboratory of Quantum Network, University of Science and Technology of China, Hefei 230026, China}
\affiliation{Hefei National Laboratory, University of Science and Technology of China, Hefei 230088, China}

\author{Guang-Can Guo}
\affiliation{Laboratory of Quantum Information, University of Science and Technology of China, Hefei 230026, China}
\affiliation{Anhui Province Key Laboratory of Quantum Network, University of Science and Technology of China, Hefei 230026, China}
\affiliation{Hefei National Laboratory, University of Science and Technology of China, Hefei 230088, China}

\author{Wei Chen}
\email{weich@ustc.edu.cn}
\affiliation{Laboratory of Quantum Information, University of Science and Technology of China, Hefei 230026, China}
\affiliation{Anhui Province Key Laboratory of Quantum Network, University of Science and Technology of China, Hefei 230026, China}
\affiliation{Hefei National Laboratory, University of Science and Technology of China, Hefei 230088, China}

\author{Zheng-Fu Han}
\affiliation{Laboratory of Quantum Information, University of Science and Technology of China, Hefei 230026, China}
\affiliation{Anhui Province Key Laboratory of Quantum Network, University of Science and Technology of China, Hefei 230026, China}
\affiliation{Hefei National Laboratory, University of Science and Technology of China, Hefei 230088, China}

\date{\today}

\begin{abstract}
Light-injection attacks pose critical security threats to quantum key distribution (QKD) systems. Conventional countermeasures, such as isolators, filters, and optical power monitoring, suffer from limited on-chip compatibility and inherent security vulnerabilities. To overcome these limitations, we propose and experimentally demonstrate an integrated \replaced[]{photorefractive-based optical power-limiting unit with attack-sensing and automatic-response capabilities}{attack sensing and automatic response unit utilizing the photorefractive effect} in a thin-film lithium niobate microring resonator. The unit provides a rejection ratio exceeding \qty{25}{dB} against non-resonant injected light. \replaced[]{Under resonant attacks with power levels above tens of microwatts, the unit autonomously attenuates the signal transmission. In our experiment, a signal attenuation of}{Under resonant attacks with power levels above tens of microwatts, the unit autonomously attenuates the signal transmission, with} \qty{14}{dB} \replaced[]{was observed under an on-chip attack power of}{attenuation measured at the maximum tested attack power of} \qty{10}{dBm}, leading to a significant suppression of the secure key rate. \added[]{We further verify its response to pulsed light injection and incorporate possible residual leakage associated with finite response time into the key-rate analysis.} This work provides a \replaced[]{sensitive, broadband, and integrated optical power-limiting approach}{highly sensitive, broadband, and fully on-chip defense mechanism} that significantly enhances the physical-layer security of QKD systems against light-injection attacks.
\end{abstract}

\maketitle

\section{Introduction}
Quantum key distribution (QKD) provides theoretically unconditional security\cite{bennett20147,lo1999unconditional}, yet its practical implementation is challenged by physical-layer attacks that exploit imperfections in real devices\cite{makarov2009controlling,lydersen2010hacking,qian2018hacking}. Light-injection attacks, where malicious light is injected into transmitters and receivers to steal information, constitute a major category including Trojan horse attacks\cite{jain2014risk,tan2021chipbased}, laser damage attacks\cite{makarov2016creation,huang2020laserdamage,ponosova2022protecting}, laser seeding attacks\cite{lovic2023quantified,sun2015effect,peng2023defending,xiao-ling2020hacking,anqi2019laserseeding}, blinding attacks\cite{lydersen2010hacking,qin2018homodyne}, and induced-photorefractive attacks\cite{ye2023induced,lu2023hacking,han2023effect}. While measurement-device-independent (MDI) QKD has successfully addressed detection-side loopholes\cite{lo2012measurement}, both MDI (source-side) and traditional protocols such as BB84 remain vulnerable to such attacks\cite{lu2023hacking}.

A key trend in advancing QKD is photonic integration, where on-chip systems offer superior miniaturization, stability, and scalability\cite{heo2025onchip,lin2025integrated}. However, conventional countermeasures against light-injection attacks face fundamental limitations in on-chip implementations. Isolators and circulators are difficult to integrate effectively and offer limited protection against attack light on the receiver side\cite{white2023integrated,yu2023integrated}. Integrated filters provide only finite rejection, and optical power monitors merely offer passive detection at high cost, resulting in poor defensive performance. These shortcomings force integrated QKD systems to rely on bulky, discrete components for defense, undermining the core benefits of integration. Moreover, such countermeasures possess inherent security vulnerabilities. For instance, isolators and circulators suffer from magnetic susceptibility and laser damage\cite{tan2022external,makarov2016creation,ponosova2022protecting}, filters lose effectiveness when attack wavelengths \replaced[]{fall}{falls} outside their spectral rejection bandwidth\cite{fadeev2025opticalpumping}, and optical power monitors are vulnerable to blinding or damage attacks\cite{lydersen2010hacking,qin2018homodyne}. \added[id=R1]{Optical power-limiting schemes provide another passive protection strategy by producing power-dependent attenuation under excessive optical illumination\cite{zhang2021securing,peng2024security}. However, existing implementations for QKD protection still face trade-offs among response threshold, insertion loss, and device size, and their on-chip integration remains challenging.} \replaced[]{Consequently, developing practical on-chip protection elements against light-injection attacks remains an important challenge for securing integrated QKD systems.}{Consequently, developing a practical, universal, and fully on-chip defense mechanism against light-injection attacks remains a critical challenge for securing QKD systems.}

Thin-film lithium niobate (TFLN) has emerged as a leading platform for integrated quantum photonics\cite{zhu2021integrated}, owing to its exceptional electro-optic performance and low propagation loss, which make it highly suitable for high-speed on-chip QKD systems\cite{heo2025onchip,lin2025integrated}. However, lithium niobate exhibits a pronounced photorefractive (PR) effect\cite{xu2021mitigating,liang2017highquality,jiang2017fast}, in which light exposure alters the material's refractive index. This effect has been identified as a potential security risk, enabling new attack vectors \cite{ye2023induced,lu2023hacking,han2023effect}. Conversely, this very property inspires a new defense strategy. \deleted[]{Here we propose to repurpose the PR effect, transforming it from a security liability into the core of an active defense scheme capable of sensing attacks and responding automatically.}\added[]{Here, we explore the possibility of repurposing the PR effect as a protective mechanism. In particular, by incorporating a microring resonator (MRR), which provides wavelength-selective coupling and intracavity field enhancement\cite{zhu2021integrated}, the PR-induced resonance shift can be used to realize power-dependent attenuation of injected light.}

\deleted[]{In this work, we present an on-chip optical fuse based on TFLN that features attack-sensing and automatic-response functions.}
\added[]{In this work, we present a photorefractive-based on-chip optical power-limiting unit on the TFLN platform for defending QKD systems against light-injection attacks. 
Unlike a sacrificial optical fuse that irreversibly interrupts an optical channel once triggered, the proposed unit provides a reversible, PR-induced power-dependent attenuation response under strong optical injection. 
In this context, the automatic response refers to a passive PR-induced process, requiring no external bias, feedback circuit, or active control signal.}
The unit combines two complementary defensive functions: spectral filtering with \added[]{a} rejection ratio \added[]{of} up to \qty{25}{dB} for non-resonant attacks and a power-sensitive, automatic-response mechanism that attenuates the signal under resonant attack powers above the microwatt level. 
We experimentally validated its defense performance in a commercial BB84 QKD system, confirming its broadband response and \replaced[]{passive automatic response}{fully automatic operation}. \added[]{We further characterize its response under pulsed light injection and provide a practical security analysis that accounts for finite response time and possible residual leakage.} 
The defense mechanism is inherently more effective against narrow-linewidth sources, making it particularly suitable for MDI QKD and continuous-variable (CV) QKD systems, which typically employ such sources\cite{shao2025highrate,li2023continuousvariable,zhang2020longdistance}. 
\replaced[]{Our proposed scheme offers a practical on-chip defense solution, featuring passive attack sensing, self-actuated response, a low response threshold, and broadband defensive coverage.}{Our proposed scheme offers a practical defense solution, featuring automatic attack sensing and response, a low response threshold, and broadband defensive coverage.} 
Most importantly, it provides a fully integrated, on-chip defense solution compatible with emerging on-chip QKD systems.

\begin{figure*}[!t]
\centering
\includegraphics[width=0.88\textwidth]{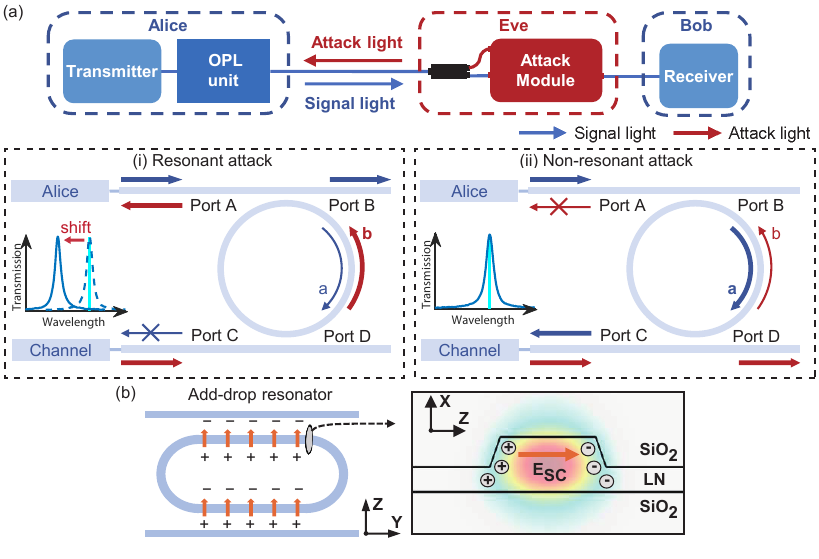}
\vspace{-0.5em}
\caption{(a) Operational principle of the \replaced[]{optical power-limiting (OPL) unit}{optical fuse} in a QKD system. The \replaced[]{OPL unit}{optical fuse} is placed at the transmitter output to protect against light-injection attacks. (\romannumeral1) Resonant attack mode: attack light at the MRR resonance wavelength induces a blue shift in the MRR spectrum via the PR effect, thereby attenuating signal transmission. (\romannumeral2) Non-resonant attack mode: the unit functions as a filter that \replaced[]{suppresses the attack light reaching the transmitter}{blocks the attack light} while maintaining normal signal light propagation. (b) Illustration of the PR effect in a TFLN waveguide, where $E_{\text{SC}}$ denotes the space-charge field.}
\vspace{-0.6em}
\label{Fig.1}
\end{figure*}

\section{\replaced[]{Mechanism of the optical power-limiting unit}{The mechanism of optical fuse}}

As illustrated in Fig.~\ref{Fig.1}(a), \replaced[]{we integrate the proposed photorefractive-based optical power-limiting unit}{a QKD system integrates our unit, the optical fuse} at the transmitter output. Under normal operation, the signal light enters through Port A and outputs to the quantum channel via Port C. The signal wavelength is aligned with the fundamental resonant wavelength of the MRR, exciting the resonance mode $a$ and enabling output through Port C. This spectral matching maximizes signal transmission, ensuring optimal QKD system operation. When Eve injects attack light, our unit operates through two distinct defense modes:

(\romannumeral1) \textbf{Resonant attack mode:} As shown in Fig.~\ref{Fig.1}(a)(\romannumeral1), when Eve injects resonant attack light into the transmitter through the quantum channel, it excites resonance mode $b$ in the MRR. The light in \added[]{the} MRR \replaced[]{induces}{triggers} the PR effect, establishing a space-charge electric field $E_{\mathrm{SC}}$ in the TFLN waveguide (Fig.~\ref{Fig.1}(b))\cite{kostritskii2009photorefractive}. Through the Pockels effect, this reduces the waveguide refractive index, producing a blue shift in the MRR resonance wavelength, \replaced[]{thereby changing}{which changing} the detuning of injected light, as described by\cite{surya2021stable,xu2021mitigating,sun2017nonlinear}:

\begin{align}
\Delta^{\prime}_{(\mathrm{a},\mathrm{b})} = \omega_{(\mathrm{a},\mathrm{b})} - \omega_{(\mathrm{sig},\mathrm{att})} + g_{\mathrm{E}}E_{\mathrm{SC}},
\label{eq:detuning}
\end{align}
where $\omega_{(\mathrm{a},\mathrm{b})} - \omega_{(\mathrm{sig},\mathrm{att})}$ represents the initial detuning between the resonance modes and input light frequencies. Within a certain range, the PR-induced detuning varies linearly with the space‑charge field $E_{\mathrm{SC}}$, with the electro‑optic coefficient $g_{\mathrm{E}}$ serving as the proportionality constant.

The signal transmission at \replaced[]{Port}{port} C is given by:
\begin{align}
\Gamma_{\text{sig,C}} = \left| \frac{\sqrt{\kappa_{\text{a,1}}\kappa_{\text{a,2}}}}{i\Delta^{\prime}_{\mathrm{a}}+\frac{\kappa_{\text{a}}}{2}} \right|^{2},
\label{eq:trans}
\end{align}
where $\kappa_{\mathrm{a},1}$ and $\kappa_{\mathrm{a},2}$ denote the external decay rates of mode $a$ at the upper and lower coupling region in the MRR, respectively. The total decay rate is given by $\kappa_{\mathrm{a}} = \kappa_{\mathrm{a},1}+\kappa_{\mathrm{a},2}+\kappa_{\mathrm{a},0}$, and $\kappa_{\mathrm{a},0}$ represents the intrinsic decay rate.

The PR-induced resonance blue shift attenuates signal transmission and consequently suppresses the secure key rate (Fig.~\ref{Fig.1}(a)(\romannumeral1)). 
\replaced[]{This effect provides a passive automatic response against resonant light-injection attacks. Under sufficiently strong injected light, the PR-induced resonance shift substantially reduces signal transmission without requiring external bias, feedback circuits, or active control signals.}{This mechanism provides an automatic response against eavesdropping by nearly ``cutting off'' the signal transmission when attack light is injected.} 
Users (Alice and Bob) can detect Eve's attacks by observing a \replaced[]{pronounced}{sharp} drop in the secure key rate and suspend the QKD system operation for security assessment (see Appendix~\ref{app:MRR} for more details).

(\romannumeral2) \textbf{Non-resonant attack mode:} When Eve injects non-resonant attack light into the MRR, the unit functions as a filter with intrinsic wavelength selectivity, providing a high rejection ratio to the attack light. Consequently, most of the attack light power is directed to Port D, \replaced[]{suppressing the injected power that can reach the transmitter}{preventing Eve from effectively injecting light into the transmitter}. This rejection ratio forces Eve to raise the attack power, which in turn induces the same PR effect and \replaced[]{activates the passive automatic response of the unit}{triggers the unit's automatic response}.

\section{Experiment}

\subsection{\replaced[]{Basic characteristics of the optical power-limiting unit}{Basic characteristics for Optical Fuse}}

\begin{figure*}[!t]
\centering\includegraphics[width=0.9\textwidth]{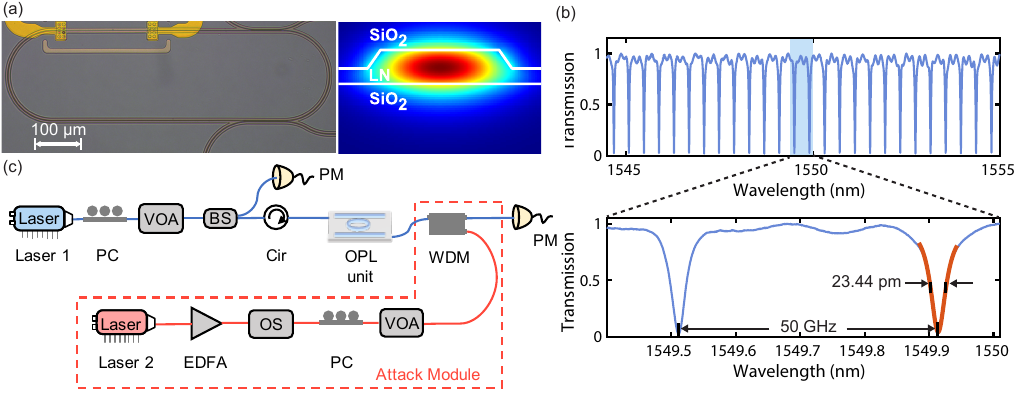}
\vspace{-0.5em}
\caption{(a) The microscope photo of the MRR and the cross-sectional schematic of the waveguide. (b) Transmission spectrum of the MRR with a zoomed-in view, showing a \replaced[]{loaded}{load} quality factor $Q_{\text{load}} = \qty{6.6e4}{}$, and an FSR of \qty{50}{GHz} for \added[]{the} unit. (c) Schematic of the experimental setup. Laser 1, a $\qty{1550}{nm}$ signal laser. PC, polarization controller. VOA, variable optical attenuator. BS, beam splitter, a $1:99$ beam splitter. PM, \added[]{optical} power meter\deleted[]{A}. Cir, a fiber circulator. \replaced[]{OPL unit, the optical power-limiting unit under test.}{Optical Fuse, the unit under test.} WDM, wavelength division multiplexer. EDFA, erbium-doped fiber amplifier. \added[id=R1]{OS, optical switch.} Laser 2, a tunable $\qty{1550}{nm}$ attack laser.}
\vspace{-0.5em}
\label{Fig.2}
\end{figure*}

Our unit is fabricated on a \qty{400}{nm}-thick x-cut TFLN-on-\ce{SiO_2} wafer with a \ce{SiO_2} cladding. The waveguides are patterned with an etch depth of \qty{200}{nm}, a top width of \qty{1}{\micro m}, and a bending radius of \qty{100}{\micro m}. Fig.~\ref{Fig.2}(a) displays the microscope photo of the unit and the cross-section schematic of the waveguide. Fig.~\ref{Fig.2}(b) exhibits a loaded quality factor of $6.6\times 10 ^{4}$ near \qty{1550}{nm} and a free spectral range (FSR) of \qty{50}{GHz}.\added[id=R1]{The unit was fiber-coupled through grating couplers, with a measured coupling loss of \qty{4.9}{dB} per grating coupler. After excluding the grating-coupler losses, the on-chip insertion loss for resonant light from Port A to Port C was estimated to be \qty{2.6}{dB}. In practical QKD deployment, these losses should be included in the system loss budget. For transmitter use, the target output intensities can be calibrated after the unit. For receiver use, this loss should be included in the receiver loss budget and accounted for in the key-rate evaluation.}

Fig.~\ref{Fig.2}(c) illustrates the experimental setup for characterizing our unit under light injection. The signal light is generated by a C-band continuous-wave (CW) laser (Laser 1) and is coupled into the unit in the fundamental transverse-electric (TE) mode. The on-chip signal power is maintained below \qty{1}{\micro W} (via VOA1) to avoid \replaced[]{inducing}{activating} \replaced[]{an unintended}{the unintended} PR effect. A wavelength division multiplexer (WDM) is used to inject the attack light and to isolate back-reflected attack light. Attack light from the tunable laser (Laser 2) is amplified by an erbium-doped fiber amplifier (EDFA), \added[id=R1]{gated by an optical switch (OS) to control the injection and interruption of the attack light,} and then adjusted in polarization and power via PC 2 and VOA 2 before being injected into Port 2 of the WDM.

\replaced[]{The following measurements characterize the unit under two representative light-injection conditions:}{The experiment is divided into two distinct conditions:}

\begin{figure*}[!t]
\centering\includegraphics{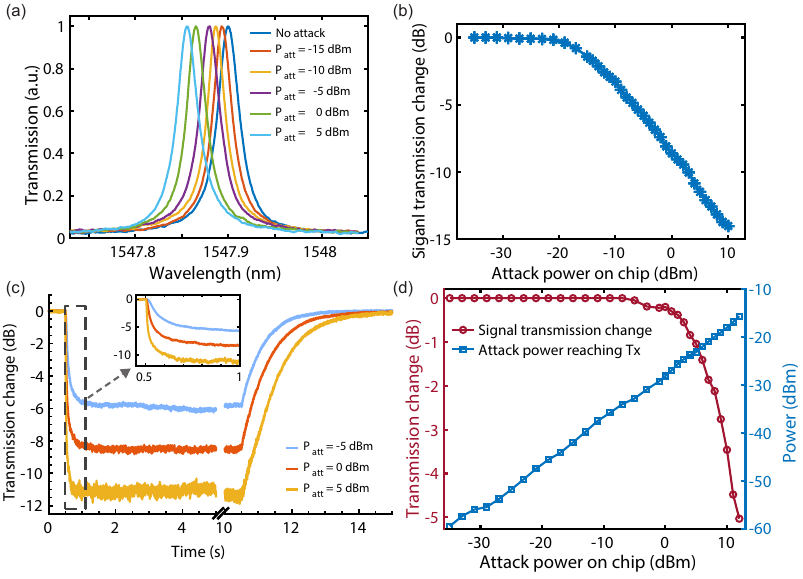}
\vspace{-0.5em}
\caption{(a) Transmission spectrum shifts under different resonant attack powers. (b) Signal transmission attenuation of \qty{1550.68}{nm} CW light under different resonant on-chip attack powers. (c) \added[]{Temporal evolution of signal transmission under different attack powers, with the attack initiated at $t = \qty{0.5}{s}$ and terminated at $t = \qty{10.5}{s}$.} (d) Signal transmission (red curve) and corresponding attack power reaching the transmitter (Tx) under different non-resonant attack powers (blue curve).}
\vspace{-0.5em}
\label{Fig.3}
\end{figure*}

(\romannumeral1) Under resonant attack conditions ($\omega_{\text{att}} = 2\pi \times \qty{193.63}{THz}\ (\qty{1548.292}{nm})$), the transmission spectrum of the unit exhibited pronounced photorefractive blue shifts, as shown in Fig.~\ref{Fig.3}(a). The blue shifts of the resonance wavelength became more pronounced with increasing attack light power, reaching a measured value of \qty{34.5}{pm} at an \added[]{on-chip} attack power of \qty{0}{dBm}. Fig.~\ref{Fig.3}(b) presents the corresponding attenuation in signal transmission \deleted[]{with}\added[]{as} the on-chip attack power was tuned from \qty{-35}{dBm} to \qty{10}{dBm}. The attenuation intensified with increasing power, becoming detectable at $P_{\text{att}}=\qty{-20}{dBm}$, and reaching \qty{14.02}{dB} at $P_{\text{att}}=\qty{10}{dBm}$. Fig.~\ref{Fig.3}(c) shows the typical dynamic response of the signal transmission. The transmission decreased rapidly upon attack initiation at \added[]{$t = \qty{0.5}{s}$}\deleted[]{t = 5 s} and recovered after the attack was terminated at \added[]{$t = \qty{10.5}{s}$}\deleted[]{t = 65 s}. \deleted[]{The unit responded within 2 s, and recovered within 2.5 s.} \added[]{The response and recovery dynamics exhibited asymmetric behavior. At an attack power of \qty{0}{dBm}, the 90\% response time was \qty{85.5}{ms}, where the response time is defined as the time required for the transmission to reach 90\% of the total attack-induced decrease. In contrast, after the attack light was removed, the 90\% recovery time was \qty{2.3}{s}, defined as the time required to recover 90\% of the same transmission change.} \replaced[]{These PR dynamics are}{This timescale of the PR effect is} consistent with the reported PR relaxation time of x-cut TFLN~\cite{sun2017nonlinear,jiang2017fast}.

(\romannumeral2) Under non-resonant attack conditions ($\omega_{\text{att}} = 2\pi \times \qty{193.65}{THz}\ (\qty{1548.091}{nm})$), Fig.~\ref{Fig.3}(d) shows the attenuation in signal transmission (red curve) and the corresponding attack power reaching the transmitter (blue curve) versus on-chip attack powers. For non-resonant attack light, the unit introduced a high rejection ratio exceeding \qty{25}{dB}, \replaced[]{strongly suppressing the attack power that could reach the transmitter while maintaining stable signal transmission}{effectively blocking eavesdropping attempts while maintaining stable signal transmission}. When the attack power exceeded approximately \qty{2}{dBm}, the finite extinction ratio of the MRR allowed the intense attack light to still induce the PR effect, thereby \replaced[]{activating}{triggering} the automatic response of the unit. For instance, at $P_{\text{att}} = \qty{5}{dBm}$, we measured a \qty{1.04}{dB} drop in signal transmission, while the attack power propagated through the unit and injected into the transmitter was \qty{-23.1}{dBm}. These results demonstrate that our unit \replaced[]{strongly rejects low-power non-resonant attacks through wavelength-selective filtering while maintaining the ability to respond automatically to high-power attacks through the PR effect}{effectively blocks low-power non-resonant attacks through a high rejection ratio while simultaneously responding to high-power attacks via the PR effect}.

\added[id=R1]{
We further examined the response of the unit under pulsed light injection. An intensity modulator (IM) was inserted after the attack laser to generate attack pulses with controlled repetition rate and duty cycle, while the signal transmission was monitored in real time. The duty cycles of \(5\%\) and \(10\%\) were chosen as representative pulsed-injection conditions. These values are also on the same order as the temporal duty cycle of short optical pulses in high-speed QKD transmitters; for example, a \qty{625}{MHz} system with \(\sim\qty{100}{ps}\) optical pulses corresponds to a duty cycle of approximately \(6\%\).
}

\begin{figure}[htbp]
\centering\includegraphics{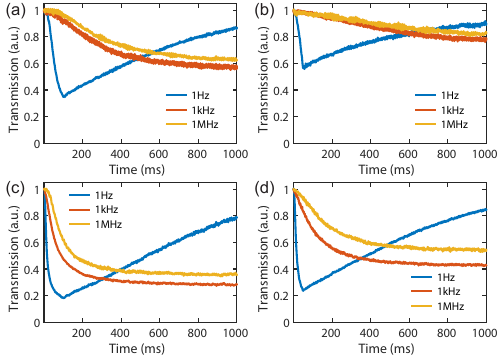}
\caption{\added[id=R1]{
Time-dependent signal transmission under pulsed attack light injection. The attack light was injected at \(t=0\) and maintained for \qty{1}{s}. 
(a) On-chip peak attack power of \qty{-5}{dBm} with a duty cycle of \(10\%\). 
(b) On-chip peak attack power of \qty{-5}{dBm} with a duty cycle of \(5\%\). 
(c) On-chip peak attack power of \qty{0}{dBm} with a duty cycle of \(10\%\). 
(d) On-chip peak attack power of \qty{0}{dBm} with a duty cycle of \(5\%\). 
In all panels, the pulse repetition rates are \qty{1}{Hz}, \qty{1}{kHz}, and \qty{1}{MHz}.
}}
\label{Fig.ex1}
\end{figure}

\added[id=R1]{
Figure~\ref{Fig.ex1}(a) shows the response to pulsed attack light with an on-chip peak power of \(-5~\mathrm{dBm}\), a duty cycle of \(10\%\), and a total injection duration of \(1~\mathrm{s}\). The peak power is specified here because pulsed injection can have a high instantaneous power even when the average power is low. At a repetition rate of \(1~\mathrm{Hz}\), corresponding to a pulse width of \(100~\mathrm{ms}\), the signal transmission decreased to approximately \(35\%\) during the attack pulse and started to recover after the pulse was switched off. For higher repetition rates of \(1~\mathrm{kHz}\) and \(1~\mathrm{MHz}\), the pulse period was shorter than the unit recovery time, so the transmission did not fully recover between adjacent pulses and instead showed a smooth decrease during the injection window. The signal transmission decreased to approximately \(57\%\) and \(63\%\) within \(1~\mathrm{s}\), respectively.
}

\added[id=R1]{
The duty-cycle dependence at the same peak power is shown in Fig.~\ref{Fig.ex1}(b). When the duty cycle was reduced from \(10\%\) to \(5\%\) at an on-chip peak power of \(-5~\mathrm{dBm}\), the induced transmission change became weaker because of the reduced average injected power and the longer relaxation interval between adjacent pulses. We then increased the on-chip peak power to \(0~\mathrm{dBm}\) to examine the peak-power dependence. As shown in Fig.~\ref{Fig.ex1}(c), a stronger transmission change was observed at the same duty cycle of \(10\%\). Figure~\ref{Fig.ex1}(d) further shows the response at the same peak power of \(0~\mathrm{dBm}\) with the duty cycle reduced to \(5\%\). Compared with the \(-5~\mathrm{dBm}\), \(5\%\)-duty-cycle case in Fig.~\ref{Fig.ex1}(b), the higher peak power produced a stronger attenuation response at the same duty cycle. These results confirm that the pulsed-injection response depends on both the peak power and the duty cycle of the attack light.
}

\added[id=R1]{
The low-repetition-rate measurement further reveals a fast component of the PR dynamics. In the \(1~\mathrm{Hz}\), \(5\%\)-duty-cycle case, the attack pulse lasted only \(50~\mathrm{ms}\), yet a substantial transmission change was already observed within this time window. This differs from the CW case, where the transmission approaches its steady state on a time scale close to seconds. Such behavior is consistent with the multi-time-scale PR response in lithium niobate, which involves multiple relaxation processes~\cite{jiang2017fast}. The response within the \(50~\mathrm{ms}\) time window may be associated with faster defect-related trapping processes. However, reaching the steady state requires not only the saturation of these fast components but also the evolution of slower defect-related relaxation components.
}

\added[id=R1]{
These measurements show that the unit remains responsive to pulsed light injection over a broad range of repetition rates and duty cycles, supporting its defensive capability against intermittent light-injection attacks.
}

\subsection{Demonstration in a commercial QKD system}

\begin{figure*}[!t]
\centering\includegraphics{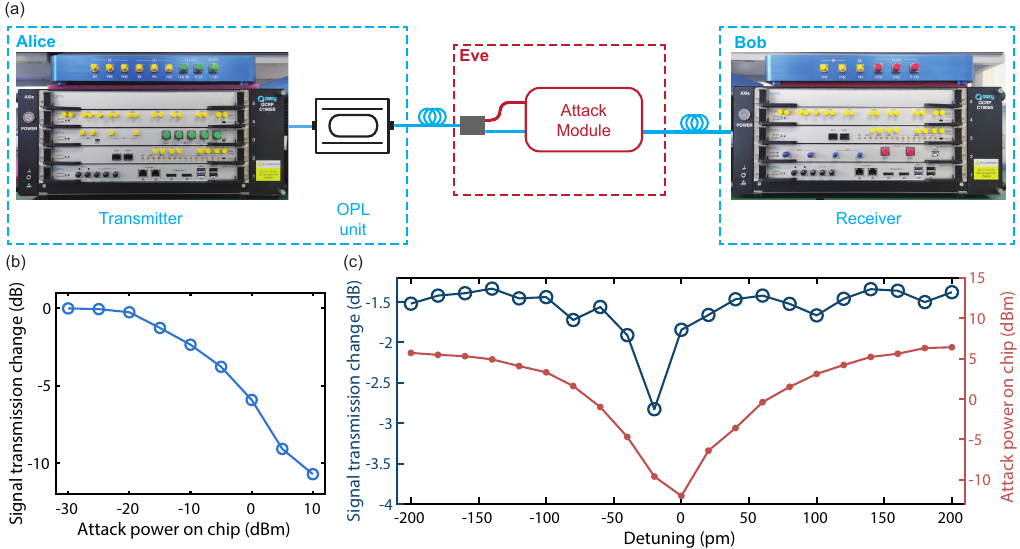}
\vspace{-0.5em}
\caption{(a) Experimental schematic for testing in a commercial QKD system. The attack module (identical to Fig.~\ref{Fig.2}(c)) intercepts the quantum channel, and the \replaced[]{optical power-limiting (OPL) unit}{optical fuse} is deployed at the transmitter output. (b) Transmission attenuation of the signal pulse under resonant attack. The width of the signal pulse is \qty{100}{ps}. (c) Signal pulse transmission attenuation (blue curve) and the on-chip attack light power (red curve) under various attack light wavelengths. The wavelength is tuned over one FSR of the unit in \qty{20}{pm} steps, centered at the resonant wavelength $\qty{1548.292}{nm}$. \replaced[]{The attack power reaching the}{The incident attack power reached} transmitter was fixed at \qty{-20}{dBm}.}
\vspace{-0.5em}
\label{Fig.4}
\end{figure*}

We further integrated the \replaced[]{optical power-limiting unit}{optical fuse} into a commercial BB84-QKD system (Qasky) (see Appendix~\ref{app:qkd_system} for more details) to evaluate its practical performance and defensive capability, as shown in Fig.~\ref{Fig.4}(a). The system consists of a transmitter (Alice), a receiver (Bob), an attack module (Eve), and the \replaced[]{unit deployed}{optical fuse which was deployed} at the transmitter output.

The QKD system was operated using a three-intensity decoy-state protocol with \qty{100}{ps} pulses at a \qty{625}{MHz} repetition rate. The intensities for the signal ($\mu$), decoy ($\nu$), and vacuum ($o$) states, along with their corresponding probabilities $P_{\lambda}$ ($\lambda \in { \mu, \nu, o }$), are summarized in Table~\ref{tab:Decoy-State Parameters}. At a transmission distance of $L = \qty{30}{km}$, the measured sifted key rate and quantum bit error rate (QBER) were \qty{4.9708e-4}{bit/pulse} and $0.0201$, respectively.

\begin{table}[t]
\caption{Decoy-state parameters.}
\begin{ruledtabular}
\begin{tabular}{cccccc}
 $\mu$ & $\nu$ & $o$ & $P_{\mu}$ & $P_{\nu}$ & $P_{o}$\\
\hline
$0.6$ & $0.2$ & $0$ & $0.8824$ & $0.0588$ & $0.0588$ \\ 
\end{tabular}
\end{ruledtabular}
\label{tab:Decoy-State Parameters}
\end{table}

To evaluate the baseline defense response of the QKD system, we first characterized the pulsed source individually. Using the attack module shown in Fig.~\ref{Fig.2}(c), we measured the signal transmission under resonant attack at $\omega_{\text{att}} = 2\pi \times \qty{193.63}{THz}\ (\qty{1548.292}{nm})$. As shown in Fig.~\ref{Fig.4}(b), the transmission exhibited a progressive reduction as the attack power increased. A measurable attenuation of \qty{0.27}{dB} was observed at $P_{\text{att}}=\qty{-20}{dBm}$, indicating a detectable response even at low power levels. This attenuation grows consistently with increasing attack power, reaching \qty{10.70}{dB} at the on-chip attack power of \qty{10}{dBm}. Such significant transmission reduction clearly predicted substantial suppression of the secure key rate. The observed attenuation was less pronounced than that measured in the earlier continuous-wave experiment. This difference arises because the defense mechanism relies on shifting the MRR resonance spectrum, \replaced[id=R1]{a process whose efficiency depends on the spectral overlap between the resonator linewidth and the signal source linewidth}{a process whose efficiency depends on both the resonator bandwidth and the signal source linewidth}. The \replaced[id=R1]{loaded linewidth}{loaded bandwidth} of our unit is approximately \qty{3}{GHz}. In contrast, the linewidth of the gain‑switched laser used here is approximately \qty{10}{GHz}, whereas that of the continuous‑wave laser is below the \unit{\MHz} level. Consequently, the defense response is inherently stronger for sources with narrower linewidths. This property makes the unit particularly suitable for protecting MDI QKD and CV QKD systems, which typically employ narrow-linewidth lasers, offering even stronger defense capability in those systems\cite{shao2025highrate,li2023continuousvariable,zhang2020longdistance}.

\begin{figure*}[!t]
\centering\includegraphics{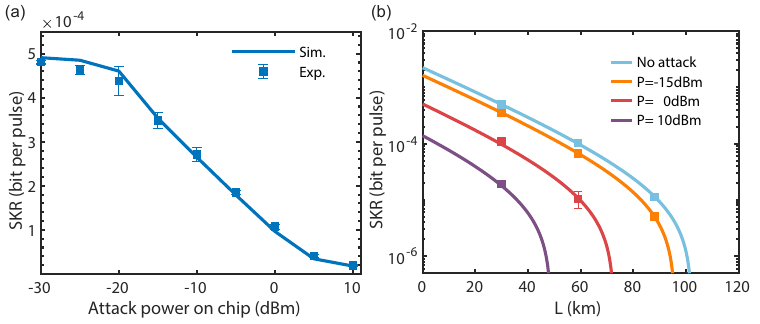}
\vspace{-0.5em}
\caption{Simulation and experimental SKR under resonant attack at different attack powers. (a) SKR versus resonant attack power at a fixed distance of \qty{30}{km}. (b) SKR versus transmission distance under different attack powers.}
\vspace{-0.5em}
\label{Fig.5}
\end{figure*}
To characterize the broadband sensing and automatic response capabilities of our unit, we measured the signal pulse transmission under light injection at various wavelengths, as shown by the blue curve in Fig.~\ref{Fig.4}(c). To evaluate the system's resilience against realistic threats, we defined a defense threshold of \qty{-20}{dBm} for the attack power reaching the transmitter. This level was chosen based on power levels reported for practical light‑injection attacks in prior studies\cite{huang2020laserdamage,ponosova2022protecting,makarov2016creation,lu2023hacking,han2023effect,fadeev2025opticalpumping,tan2021chipbased}, representing a low‑power attack condition under which the system's ability to remain secure could be meaningfully assessed. Accordingly, the attack power propagating through the \replaced[]{optical power-limiting unit}{optical fuse} and reaching \added[]{the} transmitter was fixed at this level. The attack wavelength was tuned over a span of \qty{400}{pm}, corresponding to one FSR of the MRR and centered at \added[]{the} resonant wavelength (\qty{1548.292}{nm}). The corresponding on-chip attack power injected by Eve is shown by the red curve in Fig.~\ref{Fig.4}(c).

The signal transmission exhibited significant and consistent attenuation across the entire tested wavelength range, confirming the unit's defensive capability against broadband light-injection attacks. A particularly strong transmission reduction was observed at the attack light detuning of $\Delta\lambda = \qty{-20}{pm}$. This behavior originates from the interplay between the MRR’s transmission and the PR effect. For a blue-tuned attack light, its initial detuning results in lower transmission into the MRR. However, due to the finite extinction ratio, this injected light can still induce a PR blue shift. This shift reduces the detuning of the attack light, which enables more attack light to couple into the MRR and further enhances the PR effect, resulting in a stronger transmission attenuation of the signal light. Conversely, for wavelengths longer than the resonance mode, the PR blue shift increases detuning, reducing the attack light coupling. As shown by the red curve in Fig.~\ref{Fig.4}(c), the on-chip attack power required to maintain the attack light output is higher for longer wavelengths than for shorter ones.

The spectral distribution of the on-chip attack power in Fig.~\ref{Fig.4}(c) further highlights the wavelength-selective nature of the defense. 
The non-resonant attack at $\Delta\lambda = \qty{-200}{pm}$ 
\replaced[]{requires approximately}{must increase its power by approximately} \qty{17.7}{dB} \replaced[]{higher on-chip power than}{than} 
the resonant attack. These results confirm that the unit can effectively \replaced[]{sense}{detect} and respond to attacks across a continuous spectrum, from the resonant wavelength to the non-resonant regions.

During QKD system operation, we evaluated the secure key rate (SKR) under resonant attack conditions with our unit. The experimentally measured attenuation levels from Fig.~\ref{Fig.4}(b) were incorporated into the SKR calculation model to simulate and experimentally characterize the SKR variation. Fig.~\ref{Fig.5}(a) presents both the simulated and experimental results at a transmission distance of $L = \qty{30}{km}$. The SKR began to decrease at an attack power of \qty{-20}{dBm} and declined more sharply as the power increased. At an attack power of \qty{10}{dBm}, the SKR was reduced by 96.1\%, demonstrating a strong defensive response. This suppression of the SKR confirms the effective defense response of the \replaced[]{unit}{optical fuse}.

Fig.~\ref{Fig.5}(b) further illustrates the defensive performance across various transmission distances under on-chip attack powers of \qty{-15}{dBm}, \qty{0}{dBm}, and \qty{10}{dBm}. The SKR attenuation showed a clear dependence on transmission distance. Although all distances were affected, long-haul communication exhibited disproportionately severe degradation. This pronounced SKR degradation provides a distinguishable attack signature for legitimate users (Alice and Bob), demonstrating the unit's ability to automatically sense and respond to injected light. Consequently, if Eve attempts a light-injection attack to obtain secret keys, the unit will severely attenuate the signal transmission\deleted[id = R1]{ to prevent information leakage}; Alice and Bob can then rapidly identify the abnormal drop in SKR, immediately halt system operation, and initiate a security inspection to defend against the attack.

\begin{table*}[t]
\caption{
Representative light-injection attacks demonstrated in practical QKD systems or QKD modules with reported operating conditions.
Reported powers follow the reference plane used in the corresponding study.
APD denotes avalanche photodiode, PR denotes photorefractive, and N.R. means not reported.
}
\label{tab:attack_conditions}
\centering
\scriptsize
\renewcommand{\arraystretch}{1.15}
\begin{ruledtabular}
\begin{tabular}{lllllll}
Attack class &
Target device &
Light form &
Wavelength &
Reported attack condition &
Injection time &
Reference \\
\hline

\begin{tabular}[c]{@{}l@{}}Laser-damage\\attack\end{tabular} &
\begin{tabular}[c]{@{}l@{}}Monitoring\\photodiode\end{tabular} &
CW &
\qty{1550}{nm} &
\begin{tabular}[c]{@{}l@{}}\qty{0.5}{W}--\qty{1.7}{W}\end{tabular} &
\qty{20}{s}--\qty{30}{s} &
\cite{makarov2016creation} \\

\begin{tabular}[c]{@{}l@{}}Induced PR\\attack\end{tabular} &
\begin{tabular}[c]{@{}l@{}}LiNbO$_3$\\modulator\end{tabular} &
Pulsed &
\qty{405}{nm} &
\qty{17.6}{mW} peak &
N.R. &
\cite{lu2023hacking} \\

\begin{tabular}[c]{@{}l@{}}Trojan-horse\\attack\end{tabular} &
\begin{tabular}[c]{@{}l@{}}Receiver\end{tabular} &
\begin{tabular}[c]{@{}l@{}}Pulsed\\probe light\end{tabular} &
\qty{1924}{nm} &
\begin{tabular}[c]{@{}l@{}}\qty{21.55}{\micro W} average power; \qty{5}{MHz}\end{tabular} &
N.R. &
\cite{sajeed2017invisible} \\

\begin{tabular}[c]{@{}l@{}}Detector-blinding\\attack\end{tabular} &
APD detector &
\begin{tabular}[c]{@{}l@{}}CW + trigger\\pulses\end{tabular} &
\qty{808}{nm} &
\begin{tabular}[c]{@{}l@{}}\qty{17}{\micro W} CW;\\\qty{2.6}{mW} trigger pulse\end{tabular} &
N.R. &
\cite{gerhardt2011fullfield} \\

\begin{tabular}[c]{@{}l@{}}Detector-blinding\\attack\end{tabular} &
APD detector &
\begin{tabular}[c]{@{}l@{}}CW + trigger\\pulses\end{tabular} &
\qty{1536.22}{nm} &
\begin{tabular}[c]{@{}l@{}}\qty{1.08}{mW} CW;\\\qtyrange{808}{932}{\micro W} trigger pulses\end{tabular} &
N.R. &
\cite{lydersen2010hacking} \\

\end{tabular}
\end{ruledtabular}
\end{table*}

\section{\added[]{Defense implications and residual affected-pulse analysis}}
\subsection{\added[id=R1]{Residual affected-pulse analysis}}

\added[id=R1]{
The above experiments show that the proposed unit responds to both CW and pulsed light injection. To relate these measurements to practical QKD security, we compare its operating boundary with representative light-injection attacks and incorporate the finite response time into the key-rate analysis.
}

\added[id=R1]{
Table~\ref{tab:attack_conditions} summarizes representative light-injection attacks that have been demonstrated in practical QKD systems or QKD modules with reported operating conditions. Appendix~\ref{app:practical_defense} further discusses how the measured response threshold and response time of the proposed unit relate to different attack classes. For long-pulse or sustained bright light attacks, such as laser-damage attacks, detector-blinding attacks, and induced PR attacks, the relevant attack power and injection time should be compared with the operating boundary of the unit. In many reported cases, the required optical power or exposure time lies in the regime where the unit can produce an observable signal transmission or key-rate change.
}

\added[id=R1]{
Pulsed attacks require additional consideration because a high instantaneous peak power can be hidden within a low average power. A typical example is the Trojan-horse attack, where Eve injects probe light and analyzes the back-reflected signal to obtain modulation information. Previous work has shown that such leakage can be bounded when the average energy, or equivalently the mean photon number, of the returned Trojan-horse light is sufficiently limited~\cite{lucamarini2015practical,zhang2021securing}. Therefore, the relevant security constraint is not the instantaneous peak power of each probe pulse alone, but the mean photon number of the returned light available to Eve. By limiting the injected light power entering the system, together with the internal attenuation of Alice's transmitter, our unit can help bound the returned light available to Eve and mitigate Trojan-horse attacks.
}

\added[id=R1]{
Eve may also attempt to exploit the finite response time of the unit by injecting a short burst of attack pulses and removing the attack before the unit produces an observable response. For example, in a pulsed laser-seeding attack, Eve may transiently attack the QKD transmitter within this response window~\cite{xiao-ling2020hacking,lovic2023quantified}. Such intermittent attacks cannot be completely eliminated by the unit alone. However, our pulsed injection measurements show that, for a given peak power, the unit can respond when the duty cycle or the number of injected pulses exceeds a certain level. In practical QKD, parameter estimation and key extraction are performed using measurement statistics collected over finite data blocks~\cite{tomamichel2012tight}. Over such a data block, the measured pulsed response of the unit provides an experimental bound on the number of pulses that Eve can affect without producing an observable anomaly. If Eve injects more pulses than this bound allows, the accumulated PR response produces a detectable signal transmission or key-rate change. If Eve keeps the injected pulse fraction below this bound to avoid triggering the unit response, the possible residual leakage is incorporated into the key-rate analysis.
}

\added[id=R1]{
To account for the residual leakage in QKD, we use the GLLP tagged pulse framework together with the standard decoy-state analysis~\cite{gottesman2004security,ma2005practical}. We denote by \(\Delta\) the upper bound on the fraction of emitted pulses that may be affected by Eve without producing an observable response. Over the data block used for parameter estimation, at most a fraction \(\Delta\) of the emitted pulses is therefore treated as possibly affected. Since Eve's residual injection is not conditioned on Alice's random intensity choice, we assign the same affected fraction to all intensity settings.
}

\added[id=R1]{
For each intensity setting \(a\in\{\mu,\nu,o\}\), the observed gain and error gain are written as a mixture of clean and affected contributions,}
\begin{align}
Q_a^{\mathrm{obs}}
&=
(1-\Delta)Q_a^{\mathrm{cl}}
+
\Delta Q_a^{\mathrm{atk}},\\
T_a^{\mathrm{obs}}
&=
(1-\Delta)T_a^{\mathrm{cl}}
+
\Delta T_a^{\mathrm{atk}},
\end{align}
\added[id=R1]{
where \(T_a^{\mathrm{obs}}=E_a^{\mathrm{obs}}Q_a^{\mathrm{obs}}\). Here, \(Q_a^{\mathrm{cl}}\) and \(T_a^{\mathrm{cl}}\) denote the gain and error gain of the clean (untagged) pulses, while \(Q_a^{\mathrm{atk}}\) and \(T_a^{\mathrm{atk}}\) describe the affected contribution. In the security analysis, the affected pulses are conservatively assumed to be fully controlled by Eve and may leak information to her. These pulses are therefore treated as tagged pulses and are excluded from the positive secret-key contribution. Equivalently, no secret entropy is assigned to them. The conservative intervals for the clean statistics are obtained by allowing the affected contributions to vary within the assumed bounded model, while remaining consistent with the observed statistics and the affected fraction \(\Delta\), and then taking the worst-case endpoints. We then use the worst-case clean statistics to estimate the single-photon yield and error rate, leading to a lower bound on the final key rate. In the following formulas, the superscripts \(L\) and \(U\) denote lower and upper endpoints of the corresponding conservative intervals.
}

\added[id=R1]{
The clean single-photon yield is lower bounded by}
\begin{equation}
\begin{aligned}
Y_{1,\Delta}^{L}
=
\max\Biggl\{
&\frac{\mu}{\mu\nu-\nu^2}
\Biggl[
Q_{\nu}^{\mathrm{cl},L}e^{\nu}
-
\frac{\nu^2}{\mu^2}Q_{\mu}^{\mathrm{cl},U}e^{\mu} \\
&\hspace{2.2cm}
-
\frac{\mu^2-\nu^2}{\mu^2}Q_{o}^{\mathrm{cl},U}
\Biggr],
0
\Biggr\}.
\end{aligned}
\label{eq:Y1_delta}
\end{equation}

\added[id=R1]{
The corresponding upper bound on the clean single-photon error rate is}
\begin{equation}
e_{1,\Delta}^{U}
=
\min\left\{
\frac{
T_{\nu}^{\mathrm{cl},U}e^{\nu}
-
e_0 Q_{o}^{\mathrm{cl},L}
}{
\nu Y_{1,\Delta}^{L}
},
\frac{1}{2}
\right\},
\label{eq:e1_delta}
\end{equation}
\added[id=R1]{
where \(e_0=1/2\) is the error rate of the background or vacuum contribution. The resulting lower bound on the untagged signal-state single-photon gain is}
\begin{equation}
Q_{1,\Delta}^{L}
=
(1-\Delta)\mu e^{-\mu}Y_{1,\Delta}^{L}.
\label{eq:Q1_delta}
\end{equation}

\added[id=R1]{
With these definitions, the asymptotic secure key rate is conservatively bounded by}
\begin{equation}
\begin{aligned}
R_{\Delta} \geq q \Bigl[
& -Q_{\mu}^{\mathrm{obs}} f_{\mathrm{EC}} h_2(E_{\mu}^{\mathrm{obs}}) \\
& + Q_{1,\Delta}^{L}
\left(1-h_2\left(e_{1,\Delta}^{U}\right)\right)
\Bigr],
\end{aligned}
\label{eq:decoy_GLLP_delta}
\end{equation}
\added[id=R1]{
where \(Q_{\mu}^{\mathrm{obs}}\) and \(E_{\mu}^{\mathrm{obs}}\) are the observed gain and QBER of the signal state, \(q\) is the protocol sifting factor, \(f_{\mathrm{EC}}\) is the error-correction inefficiency, and \(h_2(x)=-x\log_2 x-(1-x)\log_2(1-x)\) is the binary entropy function. When \(\Delta=0\), Eq.~\eqref{eq:decoy_GLLP_delta} reduces to the standard decoy-state GLLP formula.
}

\added[id=R1]{
This treatment is conservative. The observed statistics are first decomposed into clean and affected parts. The clean intervals are then used to estimate the worst-case single-photon yield \(Y_{1,\Delta}^{L}\) and error rate \(e_{1,\Delta}^{U}\). Only the clean single-photon contribution \(Q_{1,\Delta}^{L}\) is used as the positive secret-key contribution, while the observed signal-state gain and QBER are still used in the error-correction term because Alice and Bob cannot identify which detection events originate from clean or affected pulses during reconciliation.
}

\begin{figure}[htbp]
\centering
\includegraphics{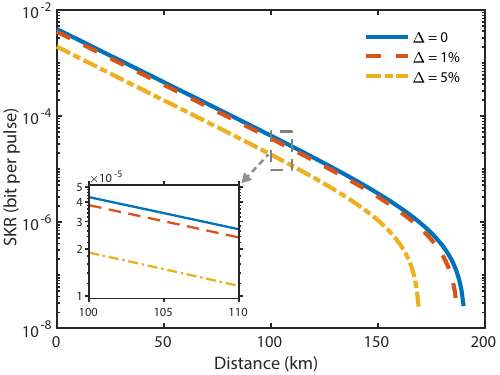}
\caption{\added[id=R1]{
Secure key rate as a function of transmission distance calculated using the modified decoy-state GLLP analysis under different bounded all-pulse affected fractions. The cases of \(\Delta=1\%\) and \(5\%\) are compared with the case without residual leakage, \(\Delta=0\). The affected pulses are conservatively treated as tagged pulses, with their intensities allowed to vary within a common single-photon-level range. The signal and weak-decoy intensities are set to \(\mu=0.6\) and \(\nu=0.2\), respectively, with a vacuum decoy state. The result illustrates that residual leakage from undetected light-injection attempts can be bounded by the measured response of the unit and included in the key-rate analysis. Other simulation parameters are as follows: dark-count probability per detector gate \(p_{\mathrm{d}}=2\times10^{-7}\), intrinsic optical error rate \(e_{\mathrm{d}}=0.25\%\), error-correction inefficiency \(f_{\mathrm{EC}}=1.25\), Bob's internal optical transmittance \(t_{\mathrm{Bob}}=0.2\), and detector efficiency \(\eta_{\mathrm{d}}=20\%\).
}}
\label{Fig.ex2}
\end{figure}

\added[id=R1]{
In Fig.~\ref{Fig.ex2}, we evaluate the influence of different bounded all-pulse affected fractions by setting \(\Delta=1\%\) and \(5\%\), and compare them with the case of \(\Delta=0\). In the key-rate simulation, the affected pulses are conservatively assumed to be fully controlled by Eve and may leak information to her. The intensities of these affected pulses are allowed to vary within a common single-photon-level range. This gives a conservative test of residual light injection, because the affected pulse intensity is allowed to vary over a broad range. In practice, this range can be further limited using experimentally characterized bounds for specific attack scenarios, which would give a higher key-rate lower bound.}

\added[id=R1]{
As expected, the key rate decreases as \(\Delta\) increases. Nevertheless, the decrease remains moderate for the considered bounded affected fractions. The choice of \(\Delta=5\%\) is motivated by our pulsed-injection experiment, where attack light with a \(5\%\) duty cycle already produced an observable response of the unit over repetition rates from \qty{1}{Hz} to \qty{1}{MHz}. This result shows that low-duty-cycle injection can already produce a measurable response under the tested conditions. Therefore, when the residual affected fraction is bounded at this level by the measured response of the unit, it can be included in the key-rate analysis with only a moderate key-rate penalty. Thus, the proposed unit can defend against light-injection attacks in two complementary ways. Strong or frequent injection produces an observable attenuation response, while residual low-duty-cycle injection can be treated as a bounded tagged-pulse contribution in the key-rate analysis.
}

\added[id=R1]{
We also considered arbitrary-polarization injection and possible MRR-induced patterning effects. The polarization extinction ratio from Port C to Port A was measured to be \qty{39.1}{dB} at the resonant wavelength, indicating strong suppression of the orthogonally polarized component. Possible patterning effects are analyzed in the low-power QKD regime, where the signal and decoy pulses are at the single-photon level and far below the PR response threshold. In this regime, the MRR operates as a passive linear filter and does not introduce intensity-dependent patterning under normal QKD operation. A detailed discussion is provided in Appendix~\ref{app:pol_patterning}.
}

\section{Discussion and conclusion}

Our defense scheme offers a versatile and practical alternative to conventional countermeasures against light-injection attacks. Conventional approaches, such as isolators, filters, and optical power monitoring, suffer from inherent limitations. Isolators are susceptible to external magnetic field manipulation\cite{tan2022external} and laser damage attacks\cite{ponosova2022protecting}. Their isolation capability degrades significantly at shorter wavelengths outside their designated operating range\cite{jain2014risk}. Filters are ineffective against attacks whose wavelengths spectrally overlap with the signal band\cite{lovic2023quantified,sun2015effect,peng2023defending,xiao-ling2020hacking,anqi2019laserseeding}. Optical power monitoring provides partial detection capability but suffers from limited spectral coverage and sensitivity degradation during laser blinding or damage attacks\cite{makarov2016creation}. Optical limiters relying on thermal effects face \replaced[]{a fundamental trade-off}{fundamental trade-off}. Bulk devices require large volumes to achieve \replaced[]{low thresholds}{low-thresholds} while introducing high insertion loss\cite{zhang2021securing}. Integrated variants exhibit \replaced[]{thresholds above the ten-milliwatt level}{over ten milliwatt-level thresholds} that limit defensive efficacy\cite{yan2014chipintegrated,alagappan2024onchip}. In contrast, our unit achieves a microwatt-level response threshold, which is four orders of magnitude lower than integrated optical limiters\cite{yan2014chipintegrated,alagappan2024onchip}, and maintains a broadband defensive response across different attack wavelengths. This combination of low threshold and broad spectral coverage enables effective protection under realistic attack conditions in QKD systems.

Our unit is well suited for advanced QKD implementations. While the bandwidth limitations of the add-drop MRR \replaced[]{require}{requires} consideration in high-speed systems, our unit has been experimentally validated in a commercial high-speed QKD system employing \qty{100}{ps} pulses at \qty{625}{MHz}. Experimental comparisons between CW and pulsed sources further demonstrate that the defensive response becomes more pronounced when the signal source has a narrower linewidth. Consequently, the proposed scheme is expected to provide superior performance in MDI-QKD and continuous-variable QKD systems, where the commonly used narrow-linewidth lasers align well with its spectral operating principle\cite{shao2025highrate,li2023continuousvariable,zhang2020longdistance}. \added[]{At the same time, the finite PR response time leaves a residual affected-pulse fraction for intermittent or low-duty-cycle pulsed attacks. As discussed above, this residual contribution can be experimentally bounded and incorporated into a decoy-state GLLP key-rate analysis as a tagged-pulse contribution.}

Material innovations offer a promising route to enhanced unit performance. For instance, \ce{Fe}:\ce{LiNbO_3} could further lower the response threshold by increasing the density of excitable charge carriers\cite{kong2012recent}. Z-cut TFLN exhibits a stronger PR effect and longer relaxation times, supporting persistent refractive index changes suitable for long-term defensive states\cite{ren2025photorefractive,jiang2017fast}.

\added[id=R1]{
The unused ports of the add-drop MRR may further serve as monitoring interfaces. Port B can monitor the through-port signal. Under normal operation, the signal wavelength is aligned to the MRR resonance and is mainly coupled to Port C, so the signal power at Port B remains at a transmission minimum. When injected light induces a PR resonance shift, the coupling of the signal light to Port C is reduced and the through-port power at Port B increases, providing an additional indication of an abnormal event. With proper calibration, continuous monitoring of Port B can also help track source power fluctuations or resonance drift during routine operation. Since the attack light is injected from Port C, Port B is not directly exposed to the main injected attack power in this configuration, which provides intrinsic protection for the monitoring detector at this port. Port D can also be used to monitor injected attack light coupled out by the MRR, although a detector placed at this port would require additional protection against direct bright-light injection. Auxiliary port monitoring can further enhance the security of the proposed defense scheme, while introducing additional components and greater system complexity.
}

In summary, we have designed and experimentally demonstrated an integrated, attack-sensing and automatic-response defense unit that utilizes the PR effect in a TFLN MRR to \replaced[]{defend QKD systems against}{against} light-injection attacks. The unit can autonomously suppress signal transmission under microwatt-level resonant attack light injection, degrading the secure key rate to alert legitimate users, while maintaining a high rejection ratio against non-resonant attacks. \added[]{The finite-response-time effect can be treated through a tagged-pulse security analysis, allowing residual leakage from undetected intermittent attacks to be bounded in the key rate evaluation.} This work transforms a material vulnerability into a defensive asset, offering a new strategy to enhance the physical-layer security of QKD systems and \replaced[]{opening new opportunities for exploiting the PR effect}{opens up new ideas for exploiting the PR effect}.

\begin{acknowledgments}
This work was funded by the Quantum Science and Technology-National Science and Technology Major Project 2021ZD0300701. This work was also supported by the Industrial Prospect and Key Core Technology Projects of Jiangsu Provincial Key R$\&$D Program (BE2022071) and National Key Laboratory of Security Communication Foundation (2023, 6142103042308).

This work was partially carried out at the USTC Center for Micro and Nanoscale Research and Fabrication.
\end{acknowledgments}
\appendix

\section{\replaced[]{Principle of the optical power-limiting unit}{Principle of Optical Fuse}}
\label{app:MRR}
\subsection{\replaced[]{Photorefractive Effect in the Optical Power-Limiting Unit}{Photorefractive Effect in the Optical Fuse}}

\begin{figure}[htbp]
    \centering
    \includegraphics[width=1\columnwidth]{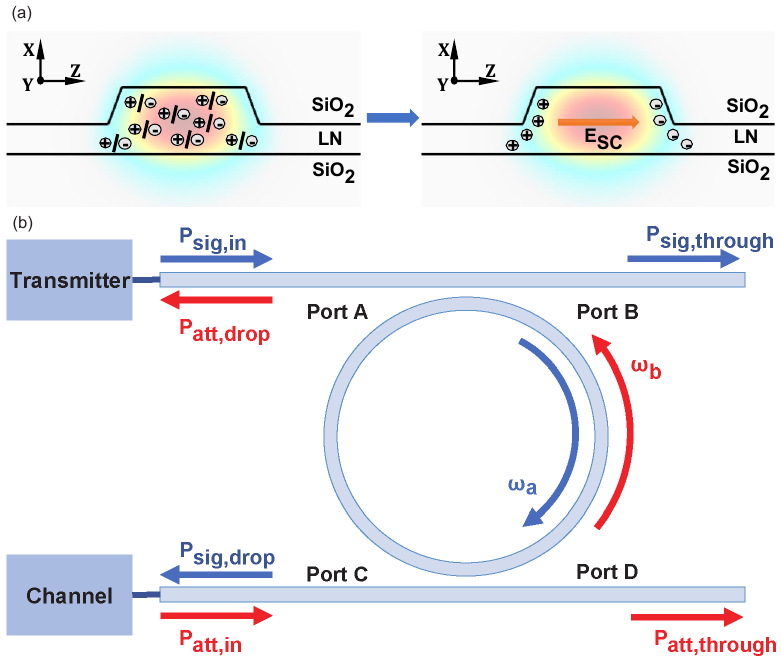}
    \vspace{-0.5em}
    \caption{(a) Schematic illustration of the PR effect in an x-cut TFLN waveguide, where $E_{\mathrm{SC}}$ denotes the space-charge electric field. (b) Structure of the \replaced[]{optical power-limiting unit}{optical fuse}. Signal light with power of $P_{\text{sig,in}}$ enters through Port A, resonates with the MRR mode $a$, and exits to the channel via Port C. Attack light injected by Eve from the channel enters through Port C. If resonant with MRR mode $b$, the attack light can propagate through the unit and reach \added[]{the} transmitter via Port A.}
    \vspace{-0.5em}
    \label{S1}
\end{figure}

The photorefractive (PR) effect refers to light-induced refractive index changes in certain materials. As illustrated in Fig.~\ref{S1}(a), in lithium niobate (LN) waveguides, incident light can excite photogenerated charge pairs from impurities and defect centers. These charges then redistribute under the influence \added[]{of} the carrier concentration distribution, the electric fields, or the photovoltaic effect, and become trapped in dark areas. This redistribution establishes a space-charge field $E_{\mathrm{SC}}$, which subsequently modifies the refractive index of the LN waveguide via the electro-optic (Pockels) effect\cite{kostritskii2009photorefractive}.

Our unit employs a microring resonator (MRR) structure that resonantly enhances the intracavity optical field, thereby increasing the total intracavity photon number $N_{\mathrm{tot}}$ to enhance the PR effect. In the MRR, the dynamics of the space-charge field $E_{\mathrm{SC}}$ are governed by\cite{surya2021stable,sun2017nonlinear}:
\begin{equation}
\frac{\mathrm{d}E_{\mathrm{SC}}}{\mathrm{d}t} = 
-\gamma_{\mathrm{e}}E_{\mathrm{SC}} + K_{\mathrm{e}} N_{\mathrm{tot}},
\label{eq:1}
\end{equation}
where $K_{\mathrm{e}}$ is the intrinsic generation coefficient of the space-charge field. $\gamma_{\mathrm{e}}$ represents the relaxation rate of the space-charge field, accounting for the decay of the PR effect over time.

\subsection{Unit Architecture and Defense Mechanism}

The proposed \replaced[]{optical power-limiting unit}{optical fuse}, shown in Fig.~\ref{S1}(b), is based on an add--drop MRR integrated at the transmitter output, connecting \added[]{the} transmitter to the quantum channel. During operation, signal light at frequency $\omega_{\text{sig}}$ enters through Port A. When resonant with MRR mode $\omega_{\text{a}}$, it couples to the quantum channel via Port C. Attack light at frequency $\omega_{\text{att}}$, injected by an eavesdropper (Eve) through Port C, can propagate through the unit and reach the transmitter if it is resonant with MRR mode $\omega_{\text{b}}$.

The system dynamics are described by the Hamiltonian\cite{surya2021stable}:
\begin{equation}
\begin{aligned}
H = &\hbar\omega_{\mathrm{a}}a^{\dagger}a + \hbar\omega_{\mathrm{b}}b^{\dagger}b \\
    &- \hbar\sqrt{\frac{\kappa_{\mathrm{a},1}P_{\mathrm{sig}}}{\hbar\omega_{\mathrm{sig}}}}(a^{\dagger}e^{-i\omega_{\mathrm{sig}}t} + ae^{i\omega_{\mathrm{sig}}t}) \\
    &- \hbar\sqrt{\frac{\kappa_{\mathrm{b},2}P_{\mathrm{att}}}{\hbar\omega_{\mathrm{att}}}}(b^{\dagger}e^{-i\omega_{\mathrm{att}}t} + be^{i\omega_{\mathrm{att}}t}),
\end{aligned}
\label{eq:2}
\end{equation}
where $P_{\mathrm{sig}}$ and $P_{\mathrm{att}}$ are the input powers of signal light and attack light, respectively. The total decay rate for each mode is given by $\kappa_{j}=\kappa_{j,1}+\kappa_{j,2}+\kappa_{j,0}$, where $j\in\{\mathrm{a},\mathrm{b}\}$. The parameters $\kappa_{j,1}$ and $\kappa_{j,2}$ denote the external decay rates of mode $j$ at the upper and lower coupling regions, respectively, while $\kappa_{j,0}$ is the intrinsic decay rate.

Incorporating both the PR effect and optical losses in the TFLN MRR, the dynamics of modes $a$ and $b$ are given by:
\begin{align}
\frac{\mathrm{d}a}{\mathrm{d}t} &= \left( -i \Delta^{\prime}_{\mathrm{a}} - \frac{\kappa_{\mathrm{a}}}{2} \right)a  + i \sqrt{\frac{\kappa_{\mathrm{a},1} P_{\mathrm{sig}}}{\hbar \omega_{\mathrm{sig}}}}, \\
\frac{\mathrm{d}b}{\mathrm{d}t} &= \left( -i \Delta^{\prime}_{\mathrm{b}} - \frac{\kappa_{\mathrm{b}}}{2} \right)b + i \sqrt{\frac{\kappa_{\mathrm{b},2} P_{\mathrm{att}}}{\hbar \omega_{\mathrm{att}}}},
\label{eq:dadt}
\end{align}
where $\Delta^{\prime}_{j}= \Delta_{j,0}+ g_{\mathrm{E}}E_{\mathrm{SC}}$ with $j\in\{\mathrm{a},\mathrm{b}\}$. Here, $\Delta_{\mathrm{a},0}=\omega_{\mathrm{a}}-\omega_{\mathrm{sig}}$ and $\Delta_{\mathrm{b},0}=\omega_{\mathrm{b}}-\omega_{\mathrm{att}}$ are the initial detunings between the resonance modes and the input light frequencies. The term $g_{\mathrm{E}}E_{\mathrm{SC}}$ represents the resonance shift induced by the PR effect, where $g_{\mathrm{E}}$ is the electro-optic coupling coefficient associated with the Pockels effect. 

Using the steady-state amplitude $a = \frac{\sqrt{\kappa_{\mathrm{a},1} P_{\mathrm{sig}}/\hbar \omega_{\mathrm{sig}}}}{i\Delta^{\prime}_{\mathrm{a}}+\frac{\kappa_{\mathrm{a}}}{2}}$, the signal transmission at Port C is obtained as:
\begin{align}
\Gamma_{\text{sig,C}} &=\left| \frac{\sqrt{\kappa_{\text{a},1}\kappa_{\text{a},2}}}{i\Delta^{\prime}_{\mathrm{a}}+\frac{\kappa_{\text{a}}}{2}} \right|^{2}.
\label{eq:transmission}
\end{align}

This formulation indicates that the PR effect, when induced by attack light, increases the detuning $\Delta^{\prime}_{\mathrm{a}}$ of the signal light, thereby reducing its transmission through Port C. Consequently, under normal operation, the signal light resonates with MRR mode $a$, maximizing transmission at Port C. When Eve injects attack light, the unit responds through two distinct defense modes:
(1) Under resonant attack, where the attack light resonates with MRR mode $b$, the high intracavity power induces the PR effect. This increases the detuning of the signal light, leading to attenuation of its transmission.
(2) Under non-resonant attack, the large detuning between the attack light and mode $b$ causes the unit to function as an optical filter. This provides a high rejection ratio against the attack light, substantially raising the power requirement of a successful attack.

\section{QKD system}
\label{app:qkd_system}

\begin{figure}[htbp]
    \centering
    \includegraphics[width=1\columnwidth]{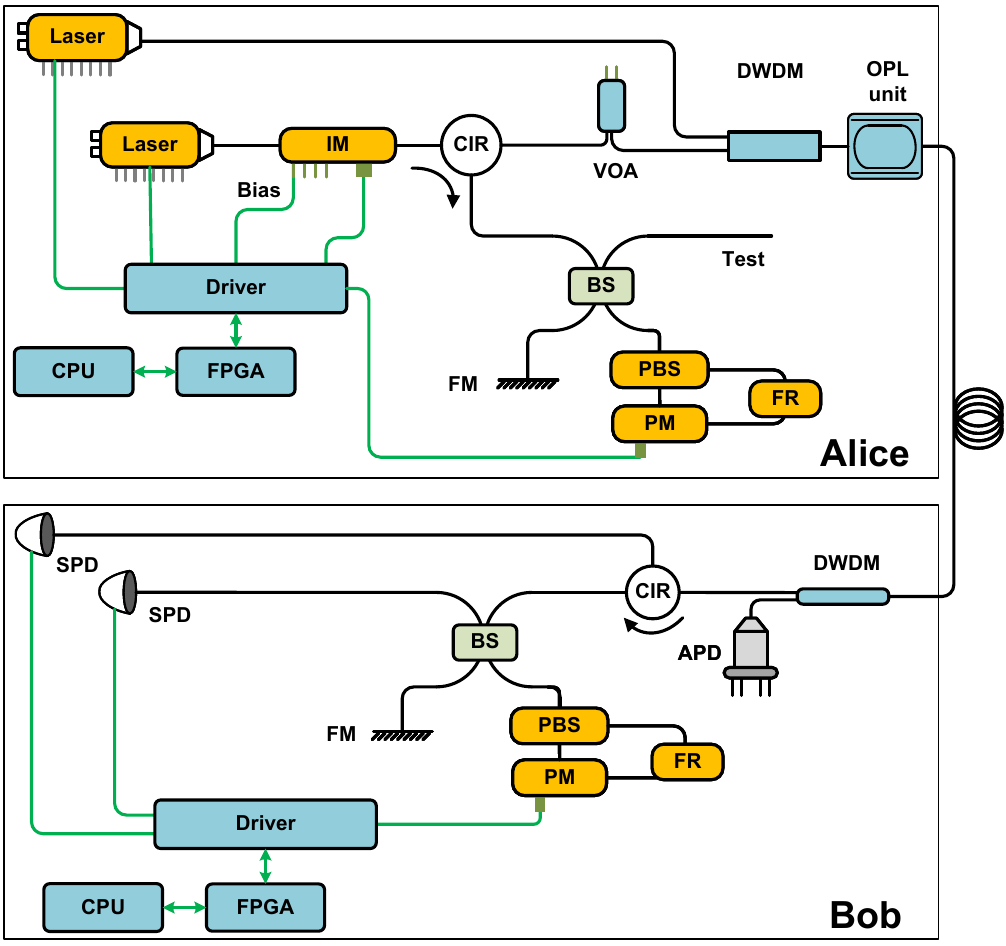}

    \caption{Schematic diagram of the phase-encoding decoy-state BB84 quantum key distribution (QKD) system. IM: intensity modulator; BS: beam splitter; PBS: polarization beam splitter; PM: phase modulator; FR: Faraday rotator; FM: Faraday mirror; VOA: variable optical attenuator; DWDM: dense wavelength division multiplexer; APD: avalanche photodiode; SPD: single-photon detector.}
 
    \label{fig:S2}
\end{figure}

\begin{figure}[htbp]
    \centering
    \includegraphics{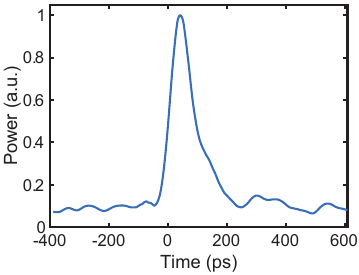}
    \caption{Waveform of the signal pulse used in the QKD system, with a measured pulse width of \qty{100}{ps}.}
    \label{fig:S3}
\end{figure}

The QKD system, implementing a phase-encoding decoy-state BB84 protocol, is depicted in Fig.~\ref{fig:S2}. On Alice's side, a gain-switched distributed feedback (DFB) laser serves as the quantum light source, emitting optical pulses at a center wavelength of \qty{1549.32}{nm} with a repetition rate of \qty{625}{MHz} and a pulse width of approximately \qty{100}{ps}. Fig.~\ref{fig:S3} shows the waveform of the signal pulse.

Alice prepares and transmits phase-randomized weak coherent states at three intensities ($\mu$, $\nu_{1}$, $\nu_{2}$). The signal ($\mu = \qty{0.6}{photon/pulse}$) and decoy ($\nu_{1} = \qty{0.2}{photon/pulse}$) states are modulated by \replaced[id=R2]{an intensity modulator (IM)}{an IM} with a bandwidth of \qty{10}{GHz}, while the vacuum decoy state ($\nu_{2}=0$) is prepared by suppressing laser emission with an extinction ratio exceeding \qty{40}{dB}.

The encoding optical module employs a Faraday-Sagnac-Michelson (FSM) interferometer, composed of \replaced[id=R2]{a beam splitter (BS), a polarization beam splitter (PBS), a phase modulator (PM), a Faraday mirror (FM) and a Faraday rotator (FR)}{a BS, a PBS, a PM, an FM and an FR}, and features inherent immunity to vibration-induced disturbances in fibers. The PM modulates the phase according to the set $\{0, \pi/2, \pi, 3\pi/2\}$ with equal probability based on random numbers.

The quantum signal is attenuated to the single-photon level using \replaced[id=R2]{a variable optical attenuator (VOA)}{a VOA} and then combined, via a dense wavelength-division multiplexer (DWDM), with a synchronization light source at \qty{1550.92}{nm}. The \replaced[id=R1]{optical power-limiting unit}{optical fuse} is placed at the transmitter output to evaluate its defense capability against light-injection attacks. Both signals then co-propagate through the optical-fiber link to the receiver (Bob).

At Bob's end, a DWDM separates the quantum and synchronization signals. The synchronization light is delivered into \replaced[id=R2]{an avalanche photodiode (APD)}{an APD}, where it is converted into an electrical clock signal that serves as the timing reference for Bob's electronic system. The quantum signal enters an FSM interferometer, identical to the one at Alice, for decoding. Two gated-mode InGaAs \replaced[id=R2]{single-photon detectors (SPDs)}{SPDs} convert optical signals into electrical pulses with a detection efficiency of \qty{20}{\%}.

The QKD system also includes a main control module and a post-processing unit, as illustrated. These consist of an electronic driver module, a timing-processing FPGA (xcku060, Xilinx), and an x86-architecture CPU (Intel Core i5-7442EQ @ \qty{2.1}{GHz}, \qty{16}{GB} RAM). Raw sampled data from the FPGA are transferred to the CPU via a PCIe 2.0 interface for post-processing. The CPU processes the raw data for basis reconciliation, error correction, and privacy amplification. Error correction is implemented using a low-density parity-check (LDPC) algorithm, achieving a tested correction rate of \qty{26.2}{Mbps} with a reconciliation efficiency of about 1.1. Privacy amplification employs a universal class of hash functions based on modular arithmetic, with an optimized multiplication algorithm for adaptive length adjustment and accelerated processing, reaching a maximum algorithmic rate of \qty{14.8}{Mbps}.

Finite-size effects are taken into account during the post-processing stage. Parameters are corrected based on the law of large numbers, and statistical fluctuations between the true values and the experimentally observed values are computed using a finite number of samples.

According to Ref.~\cite{lim2014concise}, in the finite-size scenario, the lower bound on the number of single-photon events in Z and X bases are respectively given by:

\begin{align}
    s_{1}^{X} &= \frac{\mu\tau_{1}\left\lbrack n_{\upsilon_{1}}^{XL} - n_{\upsilon_{2}}^{XU} - \frac{\upsilon_{1}^{2} - \upsilon_{2}^{2}}{\mu^{2}}\left( n_{\mu}^{XU} - \frac{s_{0}^{X}}{\tau_{0}} \right) \right\rbrack}{\mu\upsilon_{1} - \mu\upsilon_{2} - \upsilon_{1}^{2} + \upsilon_{2}^{2}},\\
    s_{1}^{Z} &= \frac{\mu\tau_{1}\left\lbrack n_{\upsilon_{1}}^{ZL} - n_{\upsilon_{2}}^{ZU} - \frac{\upsilon_{1}^{2} - \upsilon_{2}^{2}}{\mu^{2}}\left( n_{\mu}^{ZU} - \frac{s_{0}^{Z}}{\tau_{0}} \right) \right\rbrack}{\mu\upsilon_{1} - \mu\upsilon_{2} - \upsilon_{1}^{2} + \upsilon_{2}^{2}},
\end{align}

where $s_{0}^{Z}$ and $s_{0}^{X}$ denote the lower bounds on the vacuum events in the Z and X bases, respectively, and $\tau_{n}$ is the total probability of emitting an $n$-photon state. For the Z basis and each intensity $k \in \{\mu, \upsilon_{1}\}$, the finite-size bounds $n_{k}^{ZU}$ and $n_{k}^{ZL}$ (the upper and lower bounds) are derived via Hoeffding's inequality.

In addition, let $u_{1}^{X}$, $u_{1}^{Z}$ be the number of bit errors associated with the single-photon events in the X and Z bases, respectively. The upper bound on the phase error rate for the signal states at Bob's end is expressed as:

\begin{equation}
\varnothing_{z} = \frac{u_{1}^{X}}{s_{1}^{X}} + \gamma\left(\epsilon_{\text{sec}},\frac{u_{1}^{X}}{s_{1}^{X}},s_{1}^{X},s_{1}^{Z}\right).
\end{equation}
The final secure key length for a basis-sifting system under finite-size conditions is given by:
\begin{equation}
\begin{aligned}
\ell_{z} =\;& s_{0}^{Z} + s_{1}^{Z}\left( 1 - H_{2}\left( \varnothing_{z} \right) \right) \\
&- \text{leak}_{\text{EC}}
- 6\log_{2}{\frac{21}{\epsilon_{\text{sec}}}}
- \log_{2}{\frac{2}{\epsilon_{\text{cor}}}} .
\end{aligned}
\end{equation}
\replaced[]{where \(\text{leak}_{\text{EC}}\) is the information leaked during error correction, \(\epsilon_{\text{sec}}\) and \(\epsilon_{\text{cor}}\) are the secrecy and correctness parameters, and \(H_{2}(x)\) is the binary Shannon entropy function.}{where $K_{\text{cor}} = m p_{\text{sift}} Q_{\mu}$ is the total length of the output key after error correction, and $\text{leak}_{\text{EC}}$ is the information leaked during error correction, $\epsilon_{\text{sec}}$ and $\epsilon_{\text{cor}}$ are the secrecy and correctness parameters, and $H_{2}(x)$ is the binary Shannon entropy function.}

\section{\added[id=R1]{Practical defense considerations for representative light-injection attacks}}
\label{app:practical_defense}

\added[id=R1]{
To complement Table~\ref{tab:attack_conditions}, we briefly summarize how the measured operating boundary of the proposed unit relates to representative light-injection attacks. For C-band CW injection, the unit produces a measurable transmission change when the on-chip injected power exceeds approximately \(\qty{-20}{dBm}\). At an on-chip attack power of \(\qty{0}{dBm}\), the 90\% response time is below \(100~\mathrm{ms}\). Therefore, the unit should be understood as a passive protection layer against bright-light injection, rather than as an instantaneous blocker of individual optical pulses. Sustained or repeated injection can produce an observable transmission or key-rate change. Residual injection below the response boundary is treated through the bounded affected-pulse analysis.
}

\added[id=R1]{
Laser-damage attacks usually require much higher optical power and longer exposure time than the response threshold of the proposed unit. Existing demonstrations typically require hundreds of milliwatts to watt-level optical powers and second-scale or longer exposure times~\cite{huang2020laserdamage,ponosova2022protecting,makarov2016creation}. For sustained high-power attacks launched through the protected optical path, the unit response is expected to occur before the injected power and exposure time reach the reported damage regime.
}

\added[id=R1]{
For laser-seeding attacks, the relevant quantity is the residual optical power that reaches the transmitter laser diode~\cite{anqi2019laserseeding,xiao-ling2020hacking,lovic2023quantified}. The proposed unit can reduce the injected light entering the transmitter. The internal attenuation of the transmitter can further reduce the seeding power before it reaches the laser diode. Short or low-power seeding attempts may remain below the observable response boundary. Their possible contribution is therefore included in the bounded affected-pulse fraction used in the key-rate analysis.
}

\added[id=R1]{
For detector-blinding attacks, the target is usually the receiver-side single-photon detector. The required optical power depends strongly on the detector type, gating scheme, and receiver structure~\cite{makarov2009controlling,lydersen2010hacking,lydersen2010thermal,chistiakov2019controlling}. If the unit is deployed at the receiver input, sufficiently bright or repeated illumination can produce an observable transmission change. The exact protection level should be calibrated for the specific receiver implementation.
}

\added[id=R1]{
For Trojan-horse attacks, the security-relevant quantity is the mean photon number of the light returned to Eve. The proposed unit can help reduce the probe light entering the protected module. Together with the internal attenuation and isolation of the module, this reduces the returned light available to Eve and allows the corresponding leakage to be bounded using established Trojan-horse security analyses~\cite{lucamarini2015practical,zhang2021securing}.
}

\added[id=R1]{
For induced photorefraction attacks, the attacker exploits light-induced refractive-index changes in lithium-niobate modulators~\cite{lu2023hacking}. In the proposed unit, the same physical effect is intentionally used to realize the defense response. Out-of-band injected light is suppressed by the spectral response of the coupling and resonator structure. Residual injection that reaches the TFLN resonator can produce an observable attenuation when the injected power or accumulated exposure exceeds the response boundary.
}

\added[id=R1]{
Overall, the proposed unit provides two complementary forms of protection. Strong or sustained injection can produce an observable transmission or key-rate change. Residual injection below the response boundary can be experimentally bounded and incorporated into the key-rate analysis.
}
\section{\added[id=R1]{Polarization and patterning considerations}}
\label{app:pol_patterning}

\added[id=R1]{
We considered light-injection attacks with arbitrary polarization states. The unit was designed and characterized for the fundamental TE mode, and both the grating couplers and the MRR are polarization selective. For an arbitrarily polarized attack field, only the TE-mode component can be efficiently coupled into the designed resonant path, while the orthogonal polarization component is strongly attenuated. We measured the polarization extinction ratio from Port C to Port A to be \qty{39.1}{dB} at the resonant wavelength, confirming strong suppression of the orthogonally polarized component in the attack direction. For a given injected power, Eve would need to align the attack light to the TE mode to maximize the optical power coupled into the target device. Thus, TE-polarized injection represents the worst-case polarization condition for maximizing the coupled attack power and corresponds to the condition used in our CW and pulsed injection measurements.
}

\added[id=R1]{
We also considered whether the MRR could introduce intensity-dependent transmission variations between successive pulses or between different decoy-state intensities. In our QKD experiment, where the unit was placed at the transmitter output, the output signal light after the unit was set to the single-photon level. Taking into account the measured coupling and on-chip losses, the corresponding on-chip input power of the signal light was approximately \qty{-65}{dBm}, which is far below the measured PR response threshold. Therefore, normal signal and decoy pulses do not induce a measurable PR resonance shift.
}

\added[id=R1]{
In this low-power regime, the MRR acts as a passive linear filter, and its transmission does not depend on the decoy-state setting. Although the finite linewidth of the MRR can introduce wavelength-dependent filtering, this filtering is common to all decoy-state intensities because different intensity settings are generated from the same optical pulses with different attenuation levels. In addition, the loaded linewidth of the MRR is approximately \qty{3}{GHz}, corresponding to a photon lifetime on the order of tens of picoseconds according to \(\tau_{\mathrm{ph}}\approx 1/(2\pi\Delta f)\). This time scale is much shorter than the pulse period used in the QKD experiment, so the resonator does not retain appreciable optical memory between successive pulses. Therefore, the MRR does not introduce intensity-dependent patterning between successive pulses or between different decoy-state settings under normal QKD operation.
}

\bibliography{reference}

\begin{thebibliography}{51}%
\makeatletter
\providecommand \@ifxundefined [1]{%
 \@ifx{#1\undefined}
}%
\providecommand \@ifnum [1]{%
 \ifnum #1\expandafter \@firstoftwo
 \else \expandafter \@secondoftwo
 \fi
}%
\providecommand \@ifx [1]{%
 \ifx #1\expandafter \@firstoftwo
 \else \expandafter \@secondoftwo
 \fi
}%
\providecommand \natexlab [1]{#1}%
\providecommand \enquote  [1]{``#1''}%
\providecommand \bibnamefont  [1]{#1}%
\providecommand \bibfnamefont [1]{#1}%
\providecommand \citenamefont [1]{#1}%
\providecommand \href@noop [0]{\@secondoftwo}%
\providecommand \href [0]{\begingroup \@sanitize@url \@href}%
\providecommand \@href[1]{\@@startlink{#1}\@@href}%
\providecommand \@@href[1]{\endgroup#1\@@endlink}%
\providecommand \@sanitize@url [0]{\catcode `\\12\catcode `\$12\catcode `\&12\catcode `\#12\catcode `\^12\catcode `\_12\catcode `\%12\relax}%
\providecommand \@@startlink[1]{}%
\providecommand \@@endlink[0]{}%
\providecommand \url  [0]{\begingroup\@sanitize@url \@url }%
\providecommand \@url [1]{\endgroup\@href {#1}{\urlprefix }}%
\providecommand \urlprefix  [0]{URL }%
\providecommand \Eprint [0]{\href }%
\providecommand \doibase [0]{https://doi.org/}%
\providecommand \selectlanguage [0]{\@gobble}%
\providecommand \bibinfo  [0]{\@secondoftwo}%
\providecommand \bibfield  [0]{\@secondoftwo}%
\providecommand \translation [1]{[#1]}%
\providecommand \BibitemOpen [0]{}%
\providecommand \bibitemStop [0]{}%
\providecommand \bibitemNoStop [0]{.\EOS\space}%
\providecommand \EOS [0]{\spacefactor3000\relax}%
\providecommand \BibitemShut  [1]{\csname bibitem#1\endcsname}%
\let\auto@bib@innerbib\@empty
\bibitem [{\citenamefont {Bennett}\ and\ \citenamefont {Brassard}(2014)}]{bennett20147}%
  \BibitemOpen
  \bibfield  {author} {\bibinfo {author} {\bibfnamefont {C.~H.}\ \bibnamefont {Bennett}}\ and\ \bibinfo {author} {\bibfnamefont {G.}~\bibnamefont {Brassard}},\ }\bibfield  {title} {\bibinfo {title} {Quantum cryptography: Public key distribution and coin tossing},\ }\href {https://doi.org/10.1016/j.tcs.2014.05.025} {\bibfield  {journal} {\bibinfo  {journal} {Theoretical Computer Science}\ }\textbf {\bibinfo {volume} {560}},\ \bibinfo {pages} {7} (\bibinfo {year} {2014})}\BibitemShut {NoStop}%
\bibitem [{\citenamefont {Lo}\ and\ \citenamefont {Chau}(1999)}]{lo1999unconditional}%
  \BibitemOpen
  \bibfield  {author} {\bibinfo {author} {\bibfnamefont {H.-K.}\ \bibnamefont {Lo}}\ and\ \bibinfo {author} {\bibfnamefont {H.~F.}\ \bibnamefont {Chau}},\ }\bibfield  {title} {\bibinfo {title} {Unconditional security of quantum key distribution over arbitrarily long distances},\ }\href {https://doi.org/10.1126/science.283.5410.2050} {\bibfield  {journal} {\bibinfo  {journal} {Science}\ }\textbf {\bibinfo {volume} {283}},\ \bibinfo {pages} {2050} (\bibinfo {year} {1999})}\BibitemShut {NoStop}%
\bibitem [{\citenamefont {Makarov}(2009)}]{makarov2009controlling}%
  \BibitemOpen
  \bibfield  {author} {\bibinfo {author} {\bibfnamefont {V.}~\bibnamefont {Makarov}},\ }\bibfield  {title} {\bibinfo {title} {Controlling passively quenched single photon detectors by bright light},\ }\href {https://doi.org/10.1088/1367-2630/11/6/065003} {\bibfield  {journal} {\bibinfo  {journal} {New Journal of Physics}\ }\textbf {\bibinfo {volume} {11}},\ \bibinfo {pages} {065003} (\bibinfo {year} {2009})}\BibitemShut {NoStop}%
\bibitem [{\citenamefont {Lydersen}\ \emph {et~al.}(2010{\natexlab{a}})\citenamefont {Lydersen}, \citenamefont {Wiechers}, \citenamefont {Wittmann}, \citenamefont {Elser}, \citenamefont {Skaar},\ and\ \citenamefont {Makarov}}]{lydersen2010hacking}%
  \BibitemOpen
  \bibfield  {author} {\bibinfo {author} {\bibfnamefont {L.}~\bibnamefont {Lydersen}}, \bibinfo {author} {\bibfnamefont {C.}~\bibnamefont {Wiechers}}, \bibinfo {author} {\bibfnamefont {C.}~\bibnamefont {Wittmann}}, \bibinfo {author} {\bibfnamefont {D.}~\bibnamefont {Elser}}, \bibinfo {author} {\bibfnamefont {J.}~\bibnamefont {Skaar}},\ and\ \bibinfo {author} {\bibfnamefont {V.}~\bibnamefont {Makarov}},\ }\bibfield  {title} {\bibinfo {title} {Hacking commercial quantum cryptography systems by tailored bright illumination},\ }\href {https://doi.org/10.1038/nphoton.2010.214} {\bibfield  {journal} {\bibinfo  {journal} {Nature Photonics}\ }\textbf {\bibinfo {volume} {4}},\ \bibinfo {pages} {686} (\bibinfo {year} {2010}{\natexlab{a}})}\BibitemShut {NoStop}%
\bibitem [{\citenamefont {Qian}\ \emph {et~al.}(2018)\citenamefont {Qian}, \citenamefont {He}, \citenamefont {Wang}, \citenamefont {Chen}, \citenamefont {Yin}, \citenamefont {Guo},\ and\ \citenamefont {Han}}]{qian2018hacking}%
  \BibitemOpen
  \bibfield  {author} {\bibinfo {author} {\bibfnamefont {Y.-J.}\ \bibnamefont {Qian}}, \bibinfo {author} {\bibfnamefont {D.-Y.}\ \bibnamefont {He}}, \bibinfo {author} {\bibfnamefont {S.}~\bibnamefont {Wang}}, \bibinfo {author} {\bibfnamefont {W.}~\bibnamefont {Chen}}, \bibinfo {author} {\bibfnamefont {Z.-Q.}\ \bibnamefont {Yin}}, \bibinfo {author} {\bibfnamefont {G.-C.}\ \bibnamefont {Guo}},\ and\ \bibinfo {author} {\bibfnamefont {Z.-F.}\ \bibnamefont {Han}},\ }\bibfield  {title} {\bibinfo {title} {Hacking the quantum key distribution system by exploiting the avalanche-transition region of single-photon detectors},\ }\href {https://doi.org/10.1103/PhysRevApplied.10.064062} {\bibfield  {journal} {\bibinfo  {journal} {Physical Review Applied}\ }\textbf {\bibinfo {volume} {10}},\ \bibinfo {pages} {064062} (\bibinfo {year} {2018})}\BibitemShut {NoStop}%
\bibitem [{\citenamefont {Jain}\ \emph {et~al.}(2014)\citenamefont {Jain}, \citenamefont {Stiller}, \citenamefont {Khan}, \citenamefont {Makarov}, \citenamefont {Marquardt},\ and\ \citenamefont {Leuchs}}]{jain2014risk}%
  \BibitemOpen
  \bibfield  {author} {\bibinfo {author} {\bibfnamefont {N.}~\bibnamefont {Jain}}, \bibinfo {author} {\bibfnamefont {B.}~\bibnamefont {Stiller}}, \bibinfo {author} {\bibfnamefont {I.}~\bibnamefont {Khan}}, \bibinfo {author} {\bibfnamefont {V.}~\bibnamefont {Makarov}}, \bibinfo {author} {\bibfnamefont {C.}~\bibnamefont {Marquardt}},\ and\ \bibinfo {author} {\bibfnamefont {G.}~\bibnamefont {Leuchs}},\ }\bibfield  {title} {\bibinfo {title} {Risk analysis of trojan-horse attacks on practical quantum key distribution systems},\ }\href@noop {} {\bibfield  {journal} {\bibinfo  {journal} {IEEE Journal of Selected Topics in Quantum Electronics}\ }\textbf {\bibinfo {volume} {21}},\ \bibinfo {pages} {168} (\bibinfo {year} {2014})}\BibitemShut {NoStop}%
\bibitem [{\citenamefont {Tan}\ \emph {et~al.}(2021)\citenamefont {Tan}, \citenamefont {Li}, \citenamefont {Zhang}, \citenamefont {Wei},\ and\ \citenamefont {Xu}}]{tan2021chipbased}%
  \BibitemOpen
  \bibfield  {author} {\bibinfo {author} {\bibfnamefont {H.}~\bibnamefont {Tan}}, \bibinfo {author} {\bibfnamefont {W.}~\bibnamefont {Li}}, \bibinfo {author} {\bibfnamefont {L.}~\bibnamefont {Zhang}}, \bibinfo {author} {\bibfnamefont {K.}~\bibnamefont {Wei}},\ and\ \bibinfo {author} {\bibfnamefont {F.}~\bibnamefont {Xu}},\ }\bibfield  {title} {\bibinfo {title} {Chip-based quantum key distribution against trojan-horse attack},\ }\href {https://doi.org/10.1103/PhysRevApplied.15.064038} {\bibfield  {journal} {\bibinfo  {journal} {Physical Review Applied}\ }\textbf {\bibinfo {volume} {15}},\ \bibinfo {pages} {064038} (\bibinfo {year} {2021})}\BibitemShut {NoStop}%
\bibitem [{\citenamefont {Makarov}\ \emph {et~al.}(2016)\citenamefont {Makarov}, \citenamefont {Bourgoin}, \citenamefont {Chaiwongkhot}, \citenamefont {Gagn{\'e}}, \citenamefont {Jennewein}, \citenamefont {Kaiser}, \citenamefont {Kashyap}, \citenamefont {Legr{\'e}}, \citenamefont {Minshull},\ and\ \citenamefont {Sajeed}}]{makarov2016creation}%
  \BibitemOpen
  \bibfield  {author} {\bibinfo {author} {\bibfnamefont {V.}~\bibnamefont {Makarov}}, \bibinfo {author} {\bibfnamefont {J.-P.}\ \bibnamefont {Bourgoin}}, \bibinfo {author} {\bibfnamefont {P.}~\bibnamefont {Chaiwongkhot}}, \bibinfo {author} {\bibfnamefont {M.}~\bibnamefont {Gagn{\'e}}}, \bibinfo {author} {\bibfnamefont {T.}~\bibnamefont {Jennewein}}, \bibinfo {author} {\bibfnamefont {S.}~\bibnamefont {Kaiser}}, \bibinfo {author} {\bibfnamefont {R.}~\bibnamefont {Kashyap}}, \bibinfo {author} {\bibfnamefont {M.}~\bibnamefont {Legr{\'e}}}, \bibinfo {author} {\bibfnamefont {C.}~\bibnamefont {Minshull}},\ and\ \bibinfo {author} {\bibfnamefont {S.}~\bibnamefont {Sajeed}},\ }\bibfield  {title} {\bibinfo {title} {Creation of backdoors in quantum communications via laser damage},\ }\href {https://doi.org/10.1103/PhysRevA.94.030302} {\bibfield  {journal} {\bibinfo  {journal} {Physical Review A}\ }\textbf {\bibinfo {volume} {94}},\ \bibinfo {pages} {030302} (\bibinfo {year} {2016})}\BibitemShut {NoStop}%
\bibitem [{\citenamefont {Huang}\ \emph {et~al.}(2020)\citenamefont {Huang}, \citenamefont {Li}, \citenamefont {Egorov}, \citenamefont {Tchouragoulov}, \citenamefont {Kumar},\ and\ \citenamefont {Makarov}}]{huang2020laserdamage}%
  \BibitemOpen
  \bibfield  {author} {\bibinfo {author} {\bibfnamefont {A.}~\bibnamefont {Huang}}, \bibinfo {author} {\bibfnamefont {R.}~\bibnamefont {Li}}, \bibinfo {author} {\bibfnamefont {V.}~\bibnamefont {Egorov}}, \bibinfo {author} {\bibfnamefont {S.}~\bibnamefont {Tchouragoulov}}, \bibinfo {author} {\bibfnamefont {K.}~\bibnamefont {Kumar}},\ and\ \bibinfo {author} {\bibfnamefont {V.}~\bibnamefont {Makarov}},\ }\bibfield  {title} {\bibinfo {title} {Laser-damage attack against optical attenuators in quantum key distribution},\ }\href {https://doi.org/10.1103/PhysRevApplied.13.034017} {\bibfield  {journal} {\bibinfo  {journal} {Physical Review Applied}\ }\textbf {\bibinfo {volume} {13}},\ \bibinfo {pages} {034017} (\bibinfo {year} {2020})}\BibitemShut {NoStop}%
\bibitem [{\citenamefont {Ponosova}\ \emph {et~al.}(2022)\citenamefont {Ponosova}, \citenamefont {Ruzhitskaya}, \citenamefont {Chaiwongkhot}, \citenamefont {Egorov}, \citenamefont {Makarov},\ and\ \citenamefont {Huang}}]{ponosova2022protecting}%
  \BibitemOpen
  \bibfield  {author} {\bibinfo {author} {\bibfnamefont {A.}~\bibnamefont {Ponosova}}, \bibinfo {author} {\bibfnamefont {D.}~\bibnamefont {Ruzhitskaya}}, \bibinfo {author} {\bibfnamefont {P.}~\bibnamefont {Chaiwongkhot}}, \bibinfo {author} {\bibfnamefont {V.}~\bibnamefont {Egorov}}, \bibinfo {author} {\bibfnamefont {V.}~\bibnamefont {Makarov}},\ and\ \bibinfo {author} {\bibfnamefont {A.}~\bibnamefont {Huang}},\ }\bibfield  {title} {\bibinfo {title} {Protecting fiber-optic quantum key distribution sources against light-injection attacks},\ }\href {https://doi.org/10.1103/PRXQuantum.3.040307} {\bibfield  {journal} {\bibinfo  {journal} {PRX Quantum}\ }\textbf {\bibinfo {volume} {3}},\ \bibinfo {pages} {040307} (\bibinfo {year} {2022})}\BibitemShut {NoStop}%
\bibitem [{\citenamefont {Lovic}\ \emph {et~al.}(2023)\citenamefont {Lovic}, \citenamefont {Marangon}, \citenamefont {Smith}, \citenamefont {Woodward},\ and\ \citenamefont {Shields}}]{lovic2023quantified}%
  \BibitemOpen
  \bibfield  {author} {\bibinfo {author} {\bibfnamefont {V.}~\bibnamefont {Lovic}}, \bibinfo {author} {\bibfnamefont {D.}~\bibnamefont {Marangon}}, \bibinfo {author} {\bibfnamefont {P.}~\bibnamefont {Smith}}, \bibinfo {author} {\bibfnamefont {R.}~\bibnamefont {Woodward}},\ and\ \bibinfo {author} {\bibfnamefont {A.}~\bibnamefont {Shields}},\ }\bibfield  {title} {\bibinfo {title} {Quantified effects of the laser-seeding attack in quantum key distribution},\ }\href {https://doi.org/10.1103/PhysRevApplied.20.044005} {\bibfield  {journal} {\bibinfo  {journal} {Physical Review Applied}\ }\textbf {\bibinfo {volume} {20}},\ \bibinfo {pages} {044005} (\bibinfo {year} {2023})}\BibitemShut {NoStop}%
\bibitem [{\citenamefont {Sun}\ \emph {et~al.}(2015)\citenamefont {Sun}, \citenamefont {Xu}, \citenamefont {Jiang}, \citenamefont {Ma}, \citenamefont {Lo},\ and\ \citenamefont {Liang}}]{sun2015effect}%
  \BibitemOpen
  \bibfield  {author} {\bibinfo {author} {\bibfnamefont {S.-H.}\ \bibnamefont {Sun}}, \bibinfo {author} {\bibfnamefont {F.}~\bibnamefont {Xu}}, \bibinfo {author} {\bibfnamefont {M.-S.}\ \bibnamefont {Jiang}}, \bibinfo {author} {\bibfnamefont {X.-C.}\ \bibnamefont {Ma}}, \bibinfo {author} {\bibfnamefont {H.-K.}\ \bibnamefont {Lo}},\ and\ \bibinfo {author} {\bibfnamefont {L.-M.}\ \bibnamefont {Liang}},\ }\bibfield  {title} {\bibinfo {title} {Effect of source tampering in the security of quantum cryptography},\ }\href {https://doi.org/10.1103/PhysRevA.92.022304} {\bibfield  {journal} {\bibinfo  {journal} {Physical Review A}\ }\textbf {\bibinfo {volume} {92}},\ \bibinfo {pages} {022304} (\bibinfo {year} {2015})}\BibitemShut {NoStop}%
\bibitem [{\citenamefont {Peng}\ \emph {et~al.}(2023)\citenamefont {Peng}, \citenamefont {Gao}, \citenamefont {Wang}, \citenamefont {Liao}, \citenamefont {Zuo}, \citenamefont {Zhong}, \citenamefont {Huang},\ and\ \citenamefont {Guo}}]{peng2023defending}%
  \BibitemOpen
  \bibfield  {author} {\bibinfo {author} {\bibfnamefont {Q.}~\bibnamefont {Peng}}, \bibinfo {author} {\bibfnamefont {B.}~\bibnamefont {Gao}}, \bibinfo {author} {\bibfnamefont {D.}~\bibnamefont {Wang}}, \bibinfo {author} {\bibfnamefont {Q.}~\bibnamefont {Liao}}, \bibinfo {author} {\bibfnamefont {Z.}~\bibnamefont {Zuo}}, \bibinfo {author} {\bibfnamefont {H.}~\bibnamefont {Zhong}}, \bibinfo {author} {\bibfnamefont {A.}~\bibnamefont {Huang}},\ and\ \bibinfo {author} {\bibfnamefont {Y.}~\bibnamefont {Guo}},\ }\bibfield  {title} {\bibinfo {title} {Defending against a laser-seeding attack on continuous-variable quantum key distribution using an improved optical power limiter},\ }\href {https://doi.org/10.1103/PhysRevA.108.052616} {\bibfield  {journal} {\bibinfo  {journal} {Physical Review A}\ }\textbf {\bibinfo {volume} {108}},\ \bibinfo {pages} {052616} (\bibinfo {year} {2023})}\BibitemShut {NoStop}%
\bibitem [{\citenamefont {Pang}\ \emph {et~al.}(2020)\citenamefont {Pang}, \citenamefont {Yang}, \citenamefont {Zhang}, \citenamefont {Dou}, \citenamefont {Li}, \citenamefont {Gao},\ and\ \citenamefont {Jin}}]{xiao-ling2020hacking}%
  \BibitemOpen
  \bibfield  {author} {\bibinfo {author} {\bibfnamefont {X.-L.}\ \bibnamefont {Pang}}, \bibinfo {author} {\bibfnamefont {A.-L.}\ \bibnamefont {Yang}}, \bibinfo {author} {\bibfnamefont {C.-N.}\ \bibnamefont {Zhang}}, \bibinfo {author} {\bibfnamefont {J.-P.}\ \bibnamefont {Dou}}, \bibinfo {author} {\bibfnamefont {H.}~\bibnamefont {Li}}, \bibinfo {author} {\bibfnamefont {J.}~\bibnamefont {Gao}},\ and\ \bibinfo {author} {\bibfnamefont {X.-M.}\ \bibnamefont {Jin}},\ }\bibfield  {title} {\bibinfo {title} {Hacking quantum key distribution via injection locking},\ }\href {https://doi.org/10.1103/PhysRevApplied.13.034008} {\bibfield  {journal} {\bibinfo  {journal} {Phys. Rev. Appl.}\ }\textbf {\bibinfo {volume} {13}},\ \bibinfo {pages} {034008} (\bibinfo {year} {2020})}\BibitemShut {NoStop}%
\bibitem [{\citenamefont {Huang}\ \emph {et~al.}(2019)\citenamefont {Huang}, \citenamefont {Navarrete}, \citenamefont {Sun}, \citenamefont {Chaiwongkhot}, \citenamefont {Curty},\ and\ \citenamefont {Makarov}}]{anqi2019laserseeding}%
  \BibitemOpen
  \bibfield  {author} {\bibinfo {author} {\bibfnamefont {A.}~\bibnamefont {Huang}}, \bibinfo {author} {\bibfnamefont {A.}~\bibnamefont {Navarrete}}, \bibinfo {author} {\bibfnamefont {S.-H.}\ \bibnamefont {Sun}}, \bibinfo {author} {\bibfnamefont {P.}~\bibnamefont {Chaiwongkhot}}, \bibinfo {author} {\bibfnamefont {M.}~\bibnamefont {Curty}},\ and\ \bibinfo {author} {\bibfnamefont {V.}~\bibnamefont {Makarov}},\ }\bibfield  {title} {\bibinfo {title} {Laser-seeding attack in quantum key distribution},\ }\href {https://doi.org/10.1103/PhysRevApplied.12.064043} {\bibfield  {journal} {\bibinfo  {journal} {Phys. Rev. Appl.}\ }\textbf {\bibinfo {volume} {12}},\ \bibinfo {pages} {064043} (\bibinfo {year} {2019})}\BibitemShut {NoStop}%
\bibitem [{\citenamefont {Qin}\ \emph {et~al.}(2018)\citenamefont {Qin}, \citenamefont {Kumar}, \citenamefont {Makarov},\ and\ \citenamefont {All{\'e}aume}}]{qin2018homodyne}%
  \BibitemOpen
  \bibfield  {author} {\bibinfo {author} {\bibfnamefont {H.}~\bibnamefont {Qin}}, \bibinfo {author} {\bibfnamefont {R.}~\bibnamefont {Kumar}}, \bibinfo {author} {\bibfnamefont {V.}~\bibnamefont {Makarov}},\ and\ \bibinfo {author} {\bibfnamefont {R.}~\bibnamefont {All{\'e}aume}},\ }\bibfield  {title} {\bibinfo {title} {Homodyne-detector-blinding attack in continuous-variable quantum key distribution},\ }\href@noop {} {\bibfield  {journal} {\bibinfo  {journal} {Physical Review A}\ }\textbf {\bibinfo {volume} {98}},\ \bibinfo {pages} {012312} (\bibinfo {year} {2018})}\BibitemShut {NoStop}%
\bibitem [{\citenamefont {Ye}\ \emph {et~al.}(2023)\citenamefont {Ye}, \citenamefont {Chen}, \citenamefont {Zhang}, \citenamefont {Lu}, \citenamefont {Wang}, \citenamefont {Huang}, \citenamefont {Wang}, \citenamefont {He}, \citenamefont {Yin}, \citenamefont {Guo} \emph {et~al.}}]{ye2023induced}%
  \BibitemOpen
  \bibfield  {author} {\bibinfo {author} {\bibfnamefont {P.}~\bibnamefont {Ye}}, \bibinfo {author} {\bibfnamefont {W.}~\bibnamefont {Chen}}, \bibinfo {author} {\bibfnamefont {G.-W.}\ \bibnamefont {Zhang}}, \bibinfo {author} {\bibfnamefont {F.-Y.}\ \bibnamefont {Lu}}, \bibinfo {author} {\bibfnamefont {F.-X.}\ \bibnamefont {Wang}}, \bibinfo {author} {\bibfnamefont {G.-Z.}\ \bibnamefont {Huang}}, \bibinfo {author} {\bibfnamefont {S.}~\bibnamefont {Wang}}, \bibinfo {author} {\bibfnamefont {D.-Y.}\ \bibnamefont {He}}, \bibinfo {author} {\bibfnamefont {Z.-Q.}\ \bibnamefont {Yin}}, \bibinfo {author} {\bibfnamefont {G.-C.}\ \bibnamefont {Guo}}, \emph {et~al.},\ }\bibfield  {title} {\bibinfo {title} {Induced-photorefraction attack against quantum key distribution},\ }\href@noop {} {\bibfield  {journal} {\bibinfo  {journal} {Physical Review Applied}\ }\textbf {\bibinfo {volume} {19}},\ \bibinfo {pages} {054052} (\bibinfo {year} {2023})}\BibitemShut {NoStop}%
\bibitem [{\citenamefont {Lu}\ \emph {et~al.}(2023)\citenamefont {Lu}, \citenamefont {Ye}, \citenamefont {Wang}, \citenamefont {Wang}, \citenamefont {Yin}, \citenamefont {Wang}, \citenamefont {Huang}, \citenamefont {Chen}, \citenamefont {He}, \citenamefont {Fan-Yuan} \emph {et~al.}}]{lu2023hacking}%
  \BibitemOpen
  \bibfield  {author} {\bibinfo {author} {\bibfnamefont {F.-Y.}\ \bibnamefont {Lu}}, \bibinfo {author} {\bibfnamefont {P.}~\bibnamefont {Ye}}, \bibinfo {author} {\bibfnamefont {Z.-H.}\ \bibnamefont {Wang}}, \bibinfo {author} {\bibfnamefont {S.}~\bibnamefont {Wang}}, \bibinfo {author} {\bibfnamefont {Z.-Q.}\ \bibnamefont {Yin}}, \bibinfo {author} {\bibfnamefont {R.}~\bibnamefont {Wang}}, \bibinfo {author} {\bibfnamefont {X.-J.}\ \bibnamefont {Huang}}, \bibinfo {author} {\bibfnamefont {W.}~\bibnamefont {Chen}}, \bibinfo {author} {\bibfnamefont {D.-Y.}\ \bibnamefont {He}}, \bibinfo {author} {\bibfnamefont {G.-J.}\ \bibnamefont {Fan-Yuan}}, \emph {et~al.},\ }\bibfield  {title} {\bibinfo {title} {Hacking measurement-device-independent quantum key distribution},\ }\href@noop {} {\bibfield  {journal} {\bibinfo  {journal} {Optica}\ }\textbf {\bibinfo {volume} {10}},\ \bibinfo {pages} {520} (\bibinfo {year} {2023})}\BibitemShut {NoStop}%
\bibitem [{\citenamefont {Han}\ \emph {et~al.}(2023)\citenamefont {Han}, \citenamefont {Li}, \citenamefont {Tan}, \citenamefont {Zhang}, \citenamefont {Cai}, \citenamefont {Yin}, \citenamefont {Ren}, \citenamefont {Xu}, \citenamefont {Liao},\ and\ \citenamefont {Peng}}]{han2023effect}%
  \BibitemOpen
  \bibfield  {author} {\bibinfo {author} {\bibfnamefont {L.}~\bibnamefont {Han}}, \bibinfo {author} {\bibfnamefont {Y.}~\bibnamefont {Li}}, \bibinfo {author} {\bibfnamefont {H.}~\bibnamefont {Tan}}, \bibinfo {author} {\bibfnamefont {W.}~\bibnamefont {Zhang}}, \bibinfo {author} {\bibfnamefont {W.}~\bibnamefont {Cai}}, \bibinfo {author} {\bibfnamefont {J.}~\bibnamefont {Yin}}, \bibinfo {author} {\bibfnamefont {J.}~\bibnamefont {Ren}}, \bibinfo {author} {\bibfnamefont {F.}~\bibnamefont {Xu}}, \bibinfo {author} {\bibfnamefont {S.}~\bibnamefont {Liao}},\ and\ \bibinfo {author} {\bibfnamefont {C.}~\bibnamefont {Peng}},\ }\bibfield  {title} {\bibinfo {title} {Effect of light injection on the security of practical quantum key distribution},\ }\href@noop {} {\bibfield  {journal} {\bibinfo  {journal} {Physical Review Applied}\ }\textbf {\bibinfo {volume} {20}},\ \bibinfo {pages} {044013} (\bibinfo {year} {2023})}\BibitemShut {NoStop}%
\bibitem [{\citenamefont {Lo}\ \emph {et~al.}(2012)\citenamefont {Lo}, \citenamefont {Curty},\ and\ \citenamefont {Qi}}]{lo2012measurement}%
  \BibitemOpen
  \bibfield  {author} {\bibinfo {author} {\bibfnamefont {H.-K.}\ \bibnamefont {Lo}}, \bibinfo {author} {\bibfnamefont {M.}~\bibnamefont {Curty}},\ and\ \bibinfo {author} {\bibfnamefont {B.}~\bibnamefont {Qi}},\ }\bibfield  {title} {\bibinfo {title} {Measurement-device-independent quantum key distribution},\ }\href@noop {} {\bibfield  {journal} {\bibinfo  {journal} {Physical review letters}\ }\textbf {\bibinfo {volume} {108}},\ \bibinfo {pages} {130503} (\bibinfo {year} {2012})}\BibitemShut {NoStop}%
\bibitem [{\citenamefont {Heo}\ \emph {et~al.}(2025)\citenamefont {Heo}, \citenamefont {Woo}, \citenamefont {Park}, \citenamefont {Jang}, \citenamefont {Hwang}, \citenamefont {Lee}, \citenamefont {Seo}, \citenamefont {Kim}, \citenamefont {Kwon}, \citenamefont {Jung},\ and\ \citenamefont {Han}}]{heo2025onchip}%
  \BibitemOpen
  \bibfield  {author} {\bibinfo {author} {\bibfnamefont {H.}~\bibnamefont {Heo}}, \bibinfo {author} {\bibfnamefont {M.~K.}\ \bibnamefont {Woo}}, \bibinfo {author} {\bibfnamefont {C.-H.}\ \bibnamefont {Park}}, \bibinfo {author} {\bibfnamefont {H.-S.}\ \bibnamefont {Jang}}, \bibinfo {author} {\bibfnamefont {H.}~\bibnamefont {Hwang}}, \bibinfo {author} {\bibfnamefont {H.}~\bibnamefont {Lee}}, \bibinfo {author} {\bibfnamefont {M.-K.}\ \bibnamefont {Seo}}, \bibinfo {author} {\bibfnamefont {S.}~\bibnamefont {Kim}}, \bibinfo {author} {\bibfnamefont {H.}~\bibnamefont {Kwon}}, \bibinfo {author} {\bibfnamefont {H.}~\bibnamefont {Jung}},\ and\ \bibinfo {author} {\bibfnamefont {S.-W.}\ \bibnamefont {Han}},\ }\bibfield  {title} {\bibinfo {title} {On-chip quantum key distribution over field-deployed fiber using lithium niobate photonic circuit},\ }\href {https://doi.org/10.1063/5.0223694} {\bibfield  {journal} {\bibinfo  {journal} {APL Photonics}\ }\textbf {\bibinfo {volume} {10}},\ \bibinfo {pages} {031301} (\bibinfo
  {year} {2025})}\BibitemShut {NoStop}%
\bibitem [{\citenamefont {Lin}\ \emph {et~al.}(2025)\citenamefont {Lin}, \citenamefont {Gao}, \citenamefont {Zhou}, \citenamefont {Yuan}, \citenamefont {Zhu}, \citenamefont {Lin}, \citenamefont {Zhang}, \citenamefont {Huang}, \citenamefont {Cai},\ and\ \citenamefont {Yuan}}]{lin2025integrated}%
  \BibitemOpen
  \bibfield  {author} {\bibinfo {author} {\bibfnamefont {Z.}~\bibnamefont {Lin}}, \bibinfo {author} {\bibfnamefont {Y.}~\bibnamefont {Gao}}, \bibinfo {author} {\bibfnamefont {L.}~\bibnamefont {Zhou}}, \bibinfo {author} {\bibfnamefont {H.}~\bibnamefont {Yuan}}, \bibinfo {author} {\bibfnamefont {Y.}~\bibnamefont {Zhu}}, \bibinfo {author} {\bibfnamefont {Z.}~\bibnamefont {Lin}}, \bibinfo {author} {\bibfnamefont {W.}~\bibnamefont {Zhang}}, \bibinfo {author} {\bibfnamefont {Y.}~\bibnamefont {Huang}}, \bibinfo {author} {\bibfnamefont {X.-L.}\ \bibnamefont {Cai}},\ and\ \bibinfo {author} {\bibfnamefont {Z.}~\bibnamefont {Yuan}},\ }\bibfield  {title} {\bibinfo {title} {Integrated lithium niobate photonics for high-speed quantum key distribution},\ }\href {https://doi.org/10.1364/OPTICAQ.551726} {\bibfield  {journal} {\bibinfo  {journal} {Optica Quantum}\ }\textbf {\bibinfo {volume} {3}},\ \bibinfo {pages} {195} (\bibinfo {year} {2025})}\BibitemShut {NoStop}%
\bibitem [{\citenamefont {White}\ \emph {et~al.}(2023)\citenamefont {White}, \citenamefont {Ahn}, \citenamefont {Gasse}, \citenamefont {Yang}, \citenamefont {Chang}, \citenamefont {Bowers},\ and\ \citenamefont {Vu{\v c}kovi{\'c}}}]{white2023integrated}%
  \BibitemOpen
  \bibfield  {author} {\bibinfo {author} {\bibfnamefont {A.~D.}\ \bibnamefont {White}}, \bibinfo {author} {\bibfnamefont {G.~H.}\ \bibnamefont {Ahn}}, \bibinfo {author} {\bibfnamefont {K.~V.}\ \bibnamefont {Gasse}}, \bibinfo {author} {\bibfnamefont {K.~Y.}\ \bibnamefont {Yang}}, \bibinfo {author} {\bibfnamefont {L.}~\bibnamefont {Chang}}, \bibinfo {author} {\bibfnamefont {J.~E.}\ \bibnamefont {Bowers}},\ and\ \bibinfo {author} {\bibfnamefont {J.}~\bibnamefont {Vu{\v c}kovi{\'c}}},\ }\bibfield  {title} {\bibinfo {title} {Integrated passive nonlinear optical isolators},\ }\href {https://doi.org/10.1038/s41566-022-01110-y} {\bibfield  {journal} {\bibinfo  {journal} {Nature Photonics}\ }\textbf {\bibinfo {volume} {17}},\ \bibinfo {pages} {143} (\bibinfo {year} {2023})}\BibitemShut {NoStop}%
\bibitem [{\citenamefont {Yu}\ \emph {et~al.}(2023)\citenamefont {Yu}, \citenamefont {Cheng}, \citenamefont {Reimer}, \citenamefont {He}, \citenamefont {Luke}, \citenamefont {Puma}, \citenamefont {Shao}, \citenamefont {{Shams-Ansari}}, \citenamefont {Ren}, \citenamefont {Grant}, \citenamefont {Johansson}, \citenamefont {Zhang},\ and\ \citenamefont {Lon{\v c}ar}}]{yu2023integrated}%
  \BibitemOpen
  \bibfield  {author} {\bibinfo {author} {\bibfnamefont {M.}~\bibnamefont {Yu}}, \bibinfo {author} {\bibfnamefont {R.}~\bibnamefont {Cheng}}, \bibinfo {author} {\bibfnamefont {C.}~\bibnamefont {Reimer}}, \bibinfo {author} {\bibfnamefont {L.}~\bibnamefont {He}}, \bibinfo {author} {\bibfnamefont {K.}~\bibnamefont {Luke}}, \bibinfo {author} {\bibfnamefont {E.}~\bibnamefont {Puma}}, \bibinfo {author} {\bibfnamefont {L.}~\bibnamefont {Shao}}, \bibinfo {author} {\bibfnamefont {A.}~\bibnamefont {{Shams-Ansari}}}, \bibinfo {author} {\bibfnamefont {X.}~\bibnamefont {Ren}}, \bibinfo {author} {\bibfnamefont {H.~R.}\ \bibnamefont {Grant}}, \bibinfo {author} {\bibfnamefont {L.}~\bibnamefont {Johansson}}, \bibinfo {author} {\bibfnamefont {M.}~\bibnamefont {Zhang}},\ and\ \bibinfo {author} {\bibfnamefont {M.}~\bibnamefont {Lon{\v c}ar}},\ }\bibfield  {title} {\bibinfo {title} {Integrated electro-optic isolator on thin-film lithium niobate},\ }\href {https://doi.org/10.1038/s41566-023-01227-8} {\bibfield  {journal} {\bibinfo
   {journal} {Nature Photonics}\ }\textbf {\bibinfo {volume} {17}},\ \bibinfo {pages} {666} (\bibinfo {year} {2023})}\BibitemShut {NoStop}%
\bibitem [{\citenamefont {Tan}\ \emph {et~al.}(2022)\citenamefont {Tan}, \citenamefont {Zhang}, \citenamefont {Zhang}, \citenamefont {Li}, \citenamefont {Liao},\ and\ \citenamefont {Xu}}]{tan2022external}%
  \BibitemOpen
  \bibfield  {author} {\bibinfo {author} {\bibfnamefont {H.}~\bibnamefont {Tan}}, \bibinfo {author} {\bibfnamefont {W.-Y.}\ \bibnamefont {Zhang}}, \bibinfo {author} {\bibfnamefont {L.}~\bibnamefont {Zhang}}, \bibinfo {author} {\bibfnamefont {W.}~\bibnamefont {Li}}, \bibinfo {author} {\bibfnamefont {S.-K.}\ \bibnamefont {Liao}},\ and\ \bibinfo {author} {\bibfnamefont {F.}~\bibnamefont {Xu}},\ }\bibfield  {title} {\bibinfo {title} {External magnetic effect for the security of practical quantum key distribution},\ }\href@noop {} {\bibfield  {journal} {\bibinfo  {journal} {Quantum Science and Technology}\ }\textbf {\bibinfo {volume} {7}},\ \bibinfo {pages} {045008} (\bibinfo {year} {2022})}\BibitemShut {NoStop}%
\bibitem [{\citenamefont {Fadeev}\ \emph {et~al.}(2025)\citenamefont {Fadeev}, \citenamefont {Ponosova}, \citenamefont {Peng}, \citenamefont {Huang}, \citenamefont {Shakhovoy},\ and\ \citenamefont {Makarov}}]{fadeev2025opticalpumping}%
  \BibitemOpen
  \bibfield  {author} {\bibinfo {author} {\bibfnamefont {M.}~\bibnamefont {Fadeev}}, \bibinfo {author} {\bibfnamefont {A.}~\bibnamefont {Ponosova}}, \bibinfo {author} {\bibfnamefont {Q.}~\bibnamefont {Peng}}, \bibinfo {author} {\bibfnamefont {A.}~\bibnamefont {Huang}}, \bibinfo {author} {\bibfnamefont {R.}~\bibnamefont {Shakhovoy}},\ and\ \bibinfo {author} {\bibfnamefont {V.}~\bibnamefont {Makarov}},\ }\bibfield  {title} {\bibinfo {title} {Optical-pumping attack on a quantum key distribution laser source},\ }\href {https://doi.org/10.1364/OE.562406} {\bibfield  {journal} {\bibinfo  {journal} {Optics Express}\ }\textbf {\bibinfo {volume} {33}},\ \bibinfo {pages} {47001} (\bibinfo {year} {2025})}\BibitemShut {NoStop}%
\bibitem [{\citenamefont {Zhang}\ \emph {et~al.}(2021)\citenamefont {Zhang}, \citenamefont {Primaatmaja}, \citenamefont {Haw}, \citenamefont {Gong}, \citenamefont {Wang},\ and\ \citenamefont {Lim}}]{zhang2021securing}%
  \BibitemOpen
  \bibfield  {author} {\bibinfo {author} {\bibfnamefont {G.}~\bibnamefont {Zhang}}, \bibinfo {author} {\bibfnamefont {I.~W.}\ \bibnamefont {Primaatmaja}}, \bibinfo {author} {\bibfnamefont {J.~Y.}\ \bibnamefont {Haw}}, \bibinfo {author} {\bibfnamefont {X.}~\bibnamefont {Gong}}, \bibinfo {author} {\bibfnamefont {C.}~\bibnamefont {Wang}},\ and\ \bibinfo {author} {\bibfnamefont {C.~C.~W.}\ \bibnamefont {Lim}},\ }\bibfield  {title} {\bibinfo {title} {Securing practical quantum communication systems with optical power limiters},\ }\href {https://doi.org/10.1103/PRXQuantum.2.030304} {\bibfield  {journal} {\bibinfo  {journal} {PRX Quantum}\ }\textbf {\bibinfo {volume} {2}},\ \bibinfo {pages} {030304} (\bibinfo {year} {2021})}\BibitemShut {NoStop}%
\bibitem [{\citenamefont {Peng}\ \emph {et~al.}(2024)\citenamefont {Peng}, \citenamefont {Gao}, \citenamefont {Zaitsev}, \citenamefont {Wang}, \citenamefont {Ding}, \citenamefont {Liu}, \citenamefont {Liao}, \citenamefont {Guo}, \citenamefont {Huang},\ and\ \citenamefont {Wu}}]{peng2024security}%
  \BibitemOpen
  \bibfield  {author} {\bibinfo {author} {\bibfnamefont {Q.}~\bibnamefont {Peng}}, \bibinfo {author} {\bibfnamefont {B.}~\bibnamefont {Gao}}, \bibinfo {author} {\bibfnamefont {K.}~\bibnamefont {Zaitsev}}, \bibinfo {author} {\bibfnamefont {D.}~\bibnamefont {Wang}}, \bibinfo {author} {\bibfnamefont {J.}~\bibnamefont {Ding}}, \bibinfo {author} {\bibfnamefont {Y.}~\bibnamefont {Liu}}, \bibinfo {author} {\bibfnamefont {Q.}~\bibnamefont {Liao}}, \bibinfo {author} {\bibfnamefont {Y.}~\bibnamefont {Guo}}, \bibinfo {author} {\bibfnamefont {A.}~\bibnamefont {Huang}},\ and\ \bibinfo {author} {\bibfnamefont {J.}~\bibnamefont {Wu}},\ }\bibfield  {title} {\bibinfo {title} {Security boundaries of an optical-power limiter for protecting quantum-key-distribution systems},\ }\href {https://doi.org/10.1103/PhysRevApplied.21.014026} {\bibfield  {journal} {\bibinfo  {journal} {Physical Review Applied}\ }\textbf {\bibinfo {volume} {21}},\ \bibinfo {pages} {014026} (\bibinfo {year} {2024})}\BibitemShut {NoStop}%
\bibitem [{\citenamefont {Zhu}\ \emph {et~al.}(2021)\citenamefont {Zhu}, \citenamefont {Shao}, \citenamefont {Yu}, \citenamefont {Cheng}, \citenamefont {Desiatov}, \citenamefont {Xin}, \citenamefont {Hu}, \citenamefont {Holzgrafe}, \citenamefont {Ghosh}, \citenamefont {Shams-Ansari} \emph {et~al.}}]{zhu2021integrated}%
  \BibitemOpen
  \bibfield  {author} {\bibinfo {author} {\bibfnamefont {D.}~\bibnamefont {Zhu}}, \bibinfo {author} {\bibfnamefont {L.}~\bibnamefont {Shao}}, \bibinfo {author} {\bibfnamefont {M.}~\bibnamefont {Yu}}, \bibinfo {author} {\bibfnamefont {R.}~\bibnamefont {Cheng}}, \bibinfo {author} {\bibfnamefont {B.}~\bibnamefont {Desiatov}}, \bibinfo {author} {\bibfnamefont {C.}~\bibnamefont {Xin}}, \bibinfo {author} {\bibfnamefont {Y.}~\bibnamefont {Hu}}, \bibinfo {author} {\bibfnamefont {J.}~\bibnamefont {Holzgrafe}}, \bibinfo {author} {\bibfnamefont {S.}~\bibnamefont {Ghosh}}, \bibinfo {author} {\bibfnamefont {A.}~\bibnamefont {Shams-Ansari}}, \emph {et~al.},\ }\bibfield  {title} {\bibinfo {title} {Integrated photonics on thin-film lithium niobate},\ }\href@noop {} {\bibfield  {journal} {\bibinfo  {journal} {Advances in Optics and Photonics}\ }\textbf {\bibinfo {volume} {13}},\ \bibinfo {pages} {242} (\bibinfo {year} {2021})}\BibitemShut {NoStop}%
\bibitem [{\citenamefont {Xu}\ \emph {et~al.}(2021)\citenamefont {Xu}, \citenamefont {Shen}, \citenamefont {Lu}, \citenamefont {Surya}, \citenamefont {Sayem},\ and\ \citenamefont {Tang}}]{xu2021mitigating}%
  \BibitemOpen
  \bibfield  {author} {\bibinfo {author} {\bibfnamefont {Y.}~\bibnamefont {Xu}}, \bibinfo {author} {\bibfnamefont {M.}~\bibnamefont {Shen}}, \bibinfo {author} {\bibfnamefont {J.}~\bibnamefont {Lu}}, \bibinfo {author} {\bibfnamefont {J.~B.}\ \bibnamefont {Surya}}, \bibinfo {author} {\bibfnamefont {A.~A.}\ \bibnamefont {Sayem}},\ and\ \bibinfo {author} {\bibfnamefont {H.~X.}\ \bibnamefont {Tang}},\ }\bibfield  {title} {\bibinfo {title} {Mitigating photorefractive effect in thin-film lithium niobate microring resonators},\ }\href {https://doi.org/10.1364/OE.418877} {\bibfield  {journal} {\bibinfo  {journal} {Optics Express}\ }\textbf {\bibinfo {volume} {29}},\ \bibinfo {pages} {5497} (\bibinfo {year} {2021})}\BibitemShut {NoStop}%
\bibitem [{\citenamefont {Liang}\ \emph {et~al.}(2017)\citenamefont {Liang}, \citenamefont {Luo}, \citenamefont {He}, \citenamefont {Jiang},\ and\ \citenamefont {Lin}}]{liang2017highquality}%
  \BibitemOpen
  \bibfield  {author} {\bibinfo {author} {\bibfnamefont {H.}~\bibnamefont {Liang}}, \bibinfo {author} {\bibfnamefont {R.}~\bibnamefont {Luo}}, \bibinfo {author} {\bibfnamefont {Y.}~\bibnamefont {He}}, \bibinfo {author} {\bibfnamefont {H.}~\bibnamefont {Jiang}},\ and\ \bibinfo {author} {\bibfnamefont {Q.}~\bibnamefont {Lin}},\ }\bibfield  {title} {\bibinfo {title} {High-quality lithium niobate photonic crystal nanocavities},\ }\href {https://doi.org/10.1364/OPTICA.4.001251} {\bibfield  {journal} {\bibinfo  {journal} {Optica}\ }\textbf {\bibinfo {volume} {4}},\ \bibinfo {pages} {1251} (\bibinfo {year} {2017})}\BibitemShut {NoStop}%
\bibitem [{\citenamefont {Jiang}\ \emph {et~al.}(2017)\citenamefont {Jiang}, \citenamefont {Luo}, \citenamefont {Liang}, \citenamefont {Chen}, \citenamefont {Chen},\ and\ \citenamefont {Lin}}]{jiang2017fast}%
  \BibitemOpen
  \bibfield  {author} {\bibinfo {author} {\bibfnamefont {H.}~\bibnamefont {Jiang}}, \bibinfo {author} {\bibfnamefont {R.}~\bibnamefont {Luo}}, \bibinfo {author} {\bibfnamefont {H.}~\bibnamefont {Liang}}, \bibinfo {author} {\bibfnamefont {X.}~\bibnamefont {Chen}}, \bibinfo {author} {\bibfnamefont {Y.}~\bibnamefont {Chen}},\ and\ \bibinfo {author} {\bibfnamefont {Q.}~\bibnamefont {Lin}},\ }\bibfield  {title} {\bibinfo {title} {Fast response of photorefraction in lithium niobate microresonators},\ }\href {https://doi.org/10.1364/OL.42.003267} {\bibfield  {journal} {\bibinfo  {journal} {Optics Letters}\ }\textbf {\bibinfo {volume} {42}},\ \bibinfo {pages} {3267} (\bibinfo {year} {2017})}\BibitemShut {NoStop}%
\bibitem [{\citenamefont {Shao}\ \emph {et~al.}(2025)\citenamefont {Shao}, \citenamefont {Zhou}, \citenamefont {Lin}, \citenamefont {Minder}, \citenamefont {Ge}, \citenamefont {Xie}, \citenamefont {Shen}, \citenamefont {Yan}, \citenamefont {Yin},\ and\ \citenamefont {Yuan}}]{shao2025highrate}%
  \BibitemOpen
  \bibfield  {author} {\bibinfo {author} {\bibfnamefont {S.-F.}\ \bibnamefont {Shao}}, \bibinfo {author} {\bibfnamefont {L.}~\bibnamefont {Zhou}}, \bibinfo {author} {\bibfnamefont {J.}~\bibnamefont {Lin}}, \bibinfo {author} {\bibfnamefont {M.}~\bibnamefont {Minder}}, \bibinfo {author} {\bibfnamefont {C.}~\bibnamefont {Ge}}, \bibinfo {author} {\bibfnamefont {Y.-M.}\ \bibnamefont {Xie}}, \bibinfo {author} {\bibfnamefont {A.}~\bibnamefont {Shen}}, \bibinfo {author} {\bibfnamefont {Z.}~\bibnamefont {Yan}}, \bibinfo {author} {\bibfnamefont {H.-L.}\ \bibnamefont {Yin}},\ and\ \bibinfo {author} {\bibfnamefont {Z.}~\bibnamefont {Yuan}},\ }\bibfield  {title} {\bibinfo {title} {High-rate measurement-device-independent quantum communication without optical reference light},\ }\href {https://doi.org/10.1103/PhysRevX.15.021066} {\bibfield  {journal} {\bibinfo  {journal} {Physical Review X}\ }\textbf {\bibinfo {volume} {15}},\ \bibinfo {pages} {021066} (\bibinfo {year} {2025})}\BibitemShut {NoStop}%
\bibitem [{\citenamefont {Li}\ \emph {et~al.}(2023)\citenamefont {Li}, \citenamefont {Wang}, \citenamefont {Li}, \citenamefont {Huang}, \citenamefont {Guo}, \citenamefont {Lu}, \citenamefont {Zhou},\ and\ \citenamefont {Zeng}}]{li2023continuousvariable}%
  \BibitemOpen
  \bibfield  {author} {\bibinfo {author} {\bibfnamefont {L.}~\bibnamefont {Li}}, \bibinfo {author} {\bibfnamefont {T.}~\bibnamefont {Wang}}, \bibinfo {author} {\bibfnamefont {X.}~\bibnamefont {Li}}, \bibinfo {author} {\bibfnamefont {P.}~\bibnamefont {Huang}}, \bibinfo {author} {\bibfnamefont {Y.}~\bibnamefont {Guo}}, \bibinfo {author} {\bibfnamefont {L.}~\bibnamefont {Lu}}, \bibinfo {author} {\bibfnamefont {L.}~\bibnamefont {Zhou}},\ and\ \bibinfo {author} {\bibfnamefont {G.}~\bibnamefont {Zeng}},\ }\bibfield  {title} {\bibinfo {title} {Continuous-variable quantum key distribution with on-chip light sources},\ }\href {https://doi.org/10.1364/PRJ.473328} {\bibfield  {journal} {\bibinfo  {journal} {Photonics Research}\ }\textbf {\bibinfo {volume} {11}},\ \bibinfo {pages} {504} (\bibinfo {year} {2023})}\BibitemShut {NoStop}%
\bibitem [{\citenamefont {Zhang}\ \emph {et~al.}(2020)\citenamefont {Zhang}, \citenamefont {Chen}, \citenamefont {Pirandola}, \citenamefont {Wang}, \citenamefont {Zhou}, \citenamefont {Chu}, \citenamefont {Zhao}, \citenamefont {Xu}, \citenamefont {Yu},\ and\ \citenamefont {Guo}}]{zhang2020longdistance}%
  \BibitemOpen
  \bibfield  {author} {\bibinfo {author} {\bibfnamefont {Y.}~\bibnamefont {Zhang}}, \bibinfo {author} {\bibfnamefont {Z.}~\bibnamefont {Chen}}, \bibinfo {author} {\bibfnamefont {S.}~\bibnamefont {Pirandola}}, \bibinfo {author} {\bibfnamefont {X.}~\bibnamefont {Wang}}, \bibinfo {author} {\bibfnamefont {C.}~\bibnamefont {Zhou}}, \bibinfo {author} {\bibfnamefont {B.}~\bibnamefont {Chu}}, \bibinfo {author} {\bibfnamefont {Y.}~\bibnamefont {Zhao}}, \bibinfo {author} {\bibfnamefont {B.}~\bibnamefont {Xu}}, \bibinfo {author} {\bibfnamefont {S.}~\bibnamefont {Yu}},\ and\ \bibinfo {author} {\bibfnamefont {H.}~\bibnamefont {Guo}},\ }\bibfield  {title} {\bibinfo {title} {Long-distance continuous-variable quantum key distribution over 202.81 km of fiber},\ }\href {https://doi.org/10.1103/PhysRevLett.125.010502} {\bibfield  {journal} {\bibinfo  {journal} {Physical Review Letters}\ }\textbf {\bibinfo {volume} {125}},\ \bibinfo {pages} {010502} (\bibinfo {year} {2020})}\BibitemShut {NoStop}%
\bibitem [{\citenamefont {Kostritskii}(2009)}]{kostritskii2009photorefractive}%
  \BibitemOpen
  \bibfield  {author} {\bibinfo {author} {\bibfnamefont {S.~M.}\ \bibnamefont {Kostritskii}},\ }\bibfield  {title} {\bibinfo {title} {Photorefractive effect in linbo3-based integrated-optical circuits at wavelengths of third telecom window},\ }\href {https://doi.org/10.1007/s00340-009-3501-4} {\bibfield  {journal} {\bibinfo  {journal} {Applied Physics B}\ }\textbf {\bibinfo {volume} {95}},\ \bibinfo {pages} {421} (\bibinfo {year} {2009})}\BibitemShut {NoStop}%
\bibitem [{\citenamefont {Surya}\ \emph {et~al.}(2021)\citenamefont {Surya}, \citenamefont {Lu}, \citenamefont {Xu},\ and\ \citenamefont {Tang}}]{surya2021stable}%
  \BibitemOpen
  \bibfield  {author} {\bibinfo {author} {\bibfnamefont {J.~B.}\ \bibnamefont {Surya}}, \bibinfo {author} {\bibfnamefont {J.}~\bibnamefont {Lu}}, \bibinfo {author} {\bibfnamefont {Y.}~\bibnamefont {Xu}},\ and\ \bibinfo {author} {\bibfnamefont {H.~X.}\ \bibnamefont {Tang}},\ }\bibfield  {title} {\bibinfo {title} {Stable tuning of photorefractive microcavities using an auxiliary laser},\ }\href@noop {} {\bibfield  {journal} {\bibinfo  {journal} {Optics Letters}\ }\textbf {\bibinfo {volume} {46}},\ \bibinfo {pages} {328} (\bibinfo {year} {2021})}\BibitemShut {NoStop}%
\bibitem [{\citenamefont {Sun}\ \emph {et~al.}(2017)\citenamefont {Sun}, \citenamefont {Liang}, \citenamefont {Luo}, \citenamefont {Jiang}, \citenamefont {Zhang},\ and\ \citenamefont {Lin}}]{sun2017nonlinear}%
  \BibitemOpen
  \bibfield  {author} {\bibinfo {author} {\bibfnamefont {X.}~\bibnamefont {Sun}}, \bibinfo {author} {\bibfnamefont {H.}~\bibnamefont {Liang}}, \bibinfo {author} {\bibfnamefont {R.}~\bibnamefont {Luo}}, \bibinfo {author} {\bibfnamefont {W.~C.}\ \bibnamefont {Jiang}}, \bibinfo {author} {\bibfnamefont {X.-C.}\ \bibnamefont {Zhang}},\ and\ \bibinfo {author} {\bibfnamefont {Q.}~\bibnamefont {Lin}},\ }\bibfield  {title} {\bibinfo {title} {Nonlinear optical oscillation dynamics in high-q lithium niobate microresonators},\ }\href@noop {} {\bibfield  {journal} {\bibinfo  {journal} {Optics express}\ }\textbf {\bibinfo {volume} {25}},\ \bibinfo {pages} {13504} (\bibinfo {year} {2017})}\BibitemShut {NoStop}%
\bibitem [{\citenamefont {Sajeed}\ \emph {et~al.}(2017)\citenamefont {Sajeed}, \citenamefont {Minshull}, \citenamefont {Jain},\ and\ \citenamefont {Makarov}}]{sajeed2017invisible}%
  \BibitemOpen
  \bibfield  {author} {\bibinfo {author} {\bibfnamefont {S.}~\bibnamefont {Sajeed}}, \bibinfo {author} {\bibfnamefont {C.}~\bibnamefont {Minshull}}, \bibinfo {author} {\bibfnamefont {N.}~\bibnamefont {Jain}},\ and\ \bibinfo {author} {\bibfnamefont {V.}~\bibnamefont {Makarov}},\ }\bibfield  {title} {\bibinfo {title} {Invisible trojan-horse attack},\ }\href {https://doi.org/10.1038/s41598-017-08279-1} {\bibfield  {journal} {\bibinfo  {journal} {Scientific Reports}\ }\textbf {\bibinfo {volume} {7}},\ \bibinfo {pages} {8403} (\bibinfo {year} {2017})}\BibitemShut {NoStop}%
\bibitem [{\citenamefont {Gerhardt}\ \emph {et~al.}(2011)\citenamefont {Gerhardt}, \citenamefont {Liu}, \citenamefont {{Lamas-Linares}}, \citenamefont {Skaar}, \citenamefont {Kurtsiefer},\ and\ \citenamefont {Makarov}}]{gerhardt2011fullfield}%
  \BibitemOpen
  \bibfield  {author} {\bibinfo {author} {\bibfnamefont {I.}~\bibnamefont {Gerhardt}}, \bibinfo {author} {\bibfnamefont {Q.}~\bibnamefont {Liu}}, \bibinfo {author} {\bibfnamefont {A.}~\bibnamefont {{Lamas-Linares}}}, \bibinfo {author} {\bibfnamefont {J.}~\bibnamefont {Skaar}}, \bibinfo {author} {\bibfnamefont {C.}~\bibnamefont {Kurtsiefer}},\ and\ \bibinfo {author} {\bibfnamefont {V.}~\bibnamefont {Makarov}},\ }\bibfield  {title} {\bibinfo {title} {Full-field implementation of a perfect eavesdropper on a quantum cryptography system},\ }\href {https://doi.org/10.1038/ncomms1348} {\bibfield  {journal} {\bibinfo  {journal} {Nature Communications}\ }\textbf {\bibinfo {volume} {2}},\ \bibinfo {pages} {349} (\bibinfo {year} {2011})}\BibitemShut {NoStop}%
\bibitem [{\citenamefont {Lucamarini}\ \emph {et~al.}(2015)\citenamefont {Lucamarini}, \citenamefont {Choi}, \citenamefont {Ward}, \citenamefont {Dynes}, \citenamefont {Yuan},\ and\ \citenamefont {Shields}}]{lucamarini2015practical}%
  \BibitemOpen
  \bibfield  {author} {\bibinfo {author} {\bibfnamefont {M.}~\bibnamefont {Lucamarini}}, \bibinfo {author} {\bibfnamefont {I.}~\bibnamefont {Choi}}, \bibinfo {author} {\bibfnamefont {M.~B.}\ \bibnamefont {Ward}}, \bibinfo {author} {\bibfnamefont {J.~F.}\ \bibnamefont {Dynes}}, \bibinfo {author} {\bibfnamefont {Z.~L.}\ \bibnamefont {Yuan}},\ and\ \bibinfo {author} {\bibfnamefont {A.~J.}\ \bibnamefont {Shields}},\ }\bibfield  {title} {\bibinfo {title} {Practical security bounds against the trojan-horse attack in quantum key distribution},\ }\href {https://doi.org/10.1103/PhysRevX.5.031030} {\bibfield  {journal} {\bibinfo  {journal} {Physical Review X}\ }\textbf {\bibinfo {volume} {5}},\ \bibinfo {pages} {031030} (\bibinfo {year} {2015})}\BibitemShut {NoStop}%
\bibitem [{\citenamefont {Tomamichel}\ \emph {et~al.}(2012)\citenamefont {Tomamichel}, \citenamefont {Lim}, \citenamefont {Gisin},\ and\ \citenamefont {Renner}}]{tomamichel2012tight}%
  \BibitemOpen
  \bibfield  {author} {\bibinfo {author} {\bibfnamefont {M.}~\bibnamefont {Tomamichel}}, \bibinfo {author} {\bibfnamefont {C.~C.~W.}\ \bibnamefont {Lim}}, \bibinfo {author} {\bibfnamefont {N.}~\bibnamefont {Gisin}},\ and\ \bibinfo {author} {\bibfnamefont {R.}~\bibnamefont {Renner}},\ }\bibfield  {title} {\bibinfo {title} {Tight finite-key analysis for quantum cryptography},\ }\href {https://doi.org/10.1038/ncomms1631} {\bibfield  {journal} {\bibinfo  {journal} {Nature Communications}\ }\textbf {\bibinfo {volume} {3}},\ \bibinfo {pages} {634} (\bibinfo {year} {2012})}\BibitemShut {NoStop}%
\bibitem [{\citenamefont {Gottesman}\ \emph {et~al.}(2004)\citenamefont {Gottesman}, \citenamefont {Lo}, \citenamefont {L{\"u}tkenhaus},\ and\ \citenamefont {Preskill}}]{gottesman2004security}%
  \BibitemOpen
  \bibfield  {author} {\bibinfo {author} {\bibfnamefont {D.}~\bibnamefont {Gottesman}}, \bibinfo {author} {\bibfnamefont {H.-K.}\ \bibnamefont {Lo}}, \bibinfo {author} {\bibfnamefont {N.}~\bibnamefont {L{\"u}tkenhaus}},\ and\ \bibinfo {author} {\bibfnamefont {J.}~\bibnamefont {Preskill}},\ }\bibfield  {title} {\bibinfo {title} {Security of quantum key distribution with imperfect devices},\ }\href@noop {} {\bibfield  {journal} {\bibinfo  {journal} {Quantum Information \& Computation}\ }\textbf {\bibinfo {volume} {4}},\ \bibinfo {pages} {325} (\bibinfo {year} {2004})},\ \Eprint {https://arxiv.org/abs/quant-ph/0212066} {quant-ph/0212066} \BibitemShut {NoStop}%
\bibitem [{\citenamefont {Ma}\ \emph {et~al.}(2005)\citenamefont {Ma}, \citenamefont {Qi}, \citenamefont {Zhao},\ and\ \citenamefont {Lo}}]{ma2005practical}%
  \BibitemOpen
  \bibfield  {author} {\bibinfo {author} {\bibfnamefont {X.}~\bibnamefont {Ma}}, \bibinfo {author} {\bibfnamefont {B.}~\bibnamefont {Qi}}, \bibinfo {author} {\bibfnamefont {Y.}~\bibnamefont {Zhao}},\ and\ \bibinfo {author} {\bibfnamefont {H.-K.}\ \bibnamefont {Lo}},\ }\bibfield  {title} {\bibinfo {title} {Practical decoy state for quantum key distribution},\ }\href {https://doi.org/10.1103/PhysRevA.72.012326} {\bibfield  {journal} {\bibinfo  {journal} {Physical Review A}\ }\textbf {\bibinfo {volume} {72}},\ \bibinfo {pages} {012326} (\bibinfo {year} {2005})}\BibitemShut {NoStop}%
\bibitem [{\citenamefont {Yan}\ \emph {et~al.}(2014)\citenamefont {Yan}, \citenamefont {Dong}, \citenamefont {Zheng},\ and\ \citenamefont {Zhang}}]{yan2014chipintegrated}%
  \BibitemOpen
  \bibfield  {author} {\bibinfo {author} {\bibfnamefont {S.}~\bibnamefont {Yan}}, \bibinfo {author} {\bibfnamefont {J.}~\bibnamefont {Dong}}, \bibinfo {author} {\bibfnamefont {A.}~\bibnamefont {Zheng}},\ and\ \bibinfo {author} {\bibfnamefont {X.}~\bibnamefont {Zhang}},\ }\bibfield  {title} {\bibinfo {title} {Chip-integrated optical power limiter based on an all-passive micro-ring resonator},\ }\href {https://doi.org/10.1038/srep06676} {\bibfield  {journal} {\bibinfo  {journal} {Scientific Reports}\ }\textbf {\bibinfo {volume} {4}},\ \bibinfo {pages} {6676} (\bibinfo {year} {2014})}\BibitemShut {NoStop}%
\bibitem [{\citenamefont {Alagappan}\ and\ \citenamefont {Lim}(2024)}]{alagappan2024onchip}%
  \BibitemOpen
  \bibfield  {author} {\bibinfo {author} {\bibfnamefont {G.}~\bibnamefont {Alagappan}}\ and\ \bibinfo {author} {\bibfnamefont {S.~T.}\ \bibnamefont {Lim}},\ }\bibfield  {title} {\bibinfo {title} {On-chip optical power limiter for quantum communications},\ }\href {https://doi.org/10.1002/qute.202300119} {\bibfield  {journal} {\bibinfo  {journal} {Advanced Quantum Technologies}\ }\textbf {\bibinfo {volume} {7}},\ \bibinfo {pages} {2300119} (\bibinfo {year} {2024})}\BibitemShut {NoStop}%
\bibitem [{\citenamefont {Kong}\ \emph {et~al.}(2012)\citenamefont {Kong}, \citenamefont {Liu},\ and\ \citenamefont {Xu}}]{kong2012recent}%
  \BibitemOpen
  \bibfield  {author} {\bibinfo {author} {\bibfnamefont {Y.}~\bibnamefont {Kong}}, \bibinfo {author} {\bibfnamefont {S.}~\bibnamefont {Liu}},\ and\ \bibinfo {author} {\bibfnamefont {J.}~\bibnamefont {Xu}},\ }\bibfield  {title} {\bibinfo {title} {Recent advances in the photorefraction of doped lithium niobate crystals},\ }\href {https://doi.org/10.3390/ma5101954} {\bibfield  {journal} {\bibinfo  {journal} {Materials}\ }\textbf {\bibinfo {volume} {5}},\ \bibinfo {pages} {1954} (\bibinfo {year} {2012})}\BibitemShut {NoStop}%
\bibitem [{\citenamefont {Ren}\ \emph {et~al.}(2025)\citenamefont {Ren}, \citenamefont {Lee}, \citenamefont {Xue}, \citenamefont {Ou}, \citenamefont {Yu}, \citenamefont {Chen},\ and\ \citenamefont {Yu}}]{ren2025photorefractive}%
  \BibitemOpen
  \bibfield  {author} {\bibinfo {author} {\bibfnamefont {X.}~\bibnamefont {Ren}}, \bibinfo {author} {\bibfnamefont {C.-H.}\ \bibnamefont {Lee}}, \bibinfo {author} {\bibfnamefont {K.}~\bibnamefont {Xue}}, \bibinfo {author} {\bibfnamefont {S.}~\bibnamefont {Ou}}, \bibinfo {author} {\bibfnamefont {Y.}~\bibnamefont {Yu}}, \bibinfo {author} {\bibfnamefont {Z.}~\bibnamefont {Chen}},\ and\ \bibinfo {author} {\bibfnamefont {M.}~\bibnamefont {Yu}},\ }\bibfield  {title} {\bibinfo {title} {Photorefractive and pyroelectric photonic memory and long-term stability in thin-film lithium niobate microresonators},\ }\href {https://doi.org/10.1038/s44310-024-00052-3} {\bibfield  {journal} {\bibinfo  {journal} {npj Nanophotonics}\ }\textbf {\bibinfo {volume} {2}},\ \bibinfo {pages} {1} (\bibinfo {year} {2025})}\BibitemShut {NoStop}%
\bibitem [{\citenamefont {Lim}\ \emph {et~al.}(2014)\citenamefont {Lim}, \citenamefont {Curty}, \citenamefont {Walenta}, \citenamefont {Xu},\ and\ \citenamefont {Zbinden}}]{lim2014concise}%
  \BibitemOpen
  \bibfield  {author} {\bibinfo {author} {\bibfnamefont {C.~C.~W.}\ \bibnamefont {Lim}}, \bibinfo {author} {\bibfnamefont {M.}~\bibnamefont {Curty}}, \bibinfo {author} {\bibfnamefont {N.}~\bibnamefont {Walenta}}, \bibinfo {author} {\bibfnamefont {F.}~\bibnamefont {Xu}},\ and\ \bibinfo {author} {\bibfnamefont {H.}~\bibnamefont {Zbinden}},\ }\bibfield  {title} {\bibinfo {title} {Concise security bounds for practical decoy-state quantum key distribution},\ }\href@noop {} {\bibfield  {journal} {\bibinfo  {journal} {Physical Review A}\ }\textbf {\bibinfo {volume} {89}},\ \bibinfo {pages} {022307} (\bibinfo {year} {2014})}\BibitemShut {NoStop}%
\bibitem [{\citenamefont {Lydersen}\ \emph {et~al.}(2010{\natexlab{b}})\citenamefont {Lydersen}, \citenamefont {Wiechers}, \citenamefont {Wittmann}, \citenamefont {Elser}, \citenamefont {Skaar},\ and\ \citenamefont {Makarov}}]{lydersen2010thermal}%
  \BibitemOpen
  \bibfield  {author} {\bibinfo {author} {\bibfnamefont {L.}~\bibnamefont {Lydersen}}, \bibinfo {author} {\bibfnamefont {C.}~\bibnamefont {Wiechers}}, \bibinfo {author} {\bibfnamefont {C.}~\bibnamefont {Wittmann}}, \bibinfo {author} {\bibfnamefont {D.}~\bibnamefont {Elser}}, \bibinfo {author} {\bibfnamefont {J.}~\bibnamefont {Skaar}},\ and\ \bibinfo {author} {\bibfnamefont {V.}~\bibnamefont {Makarov}},\ }\bibfield  {title} {\bibinfo {title} {Thermal blinding of gated detectors in quantum cryptography},\ }\href {https://doi.org/10.1364/OE.18.027938} {\bibfield  {journal} {\bibinfo  {journal} {Optics Express}\ }\textbf {\bibinfo {volume} {18}},\ \bibinfo {pages} {27938} (\bibinfo {year} {2010}{\natexlab{b}})}\BibitemShut {NoStop}%
\bibitem [{\citenamefont {Chistiakov}\ \emph {et~al.}(2019)\citenamefont {Chistiakov}, \citenamefont {Huang}, \citenamefont {Egorov},\ and\ \citenamefont {Makarov}}]{chistiakov2019controlling}%
  \BibitemOpen
  \bibfield  {author} {\bibinfo {author} {\bibfnamefont {V.}~\bibnamefont {Chistiakov}}, \bibinfo {author} {\bibfnamefont {A.}~\bibnamefont {Huang}}, \bibinfo {author} {\bibfnamefont {V.}~\bibnamefont {Egorov}},\ and\ \bibinfo {author} {\bibfnamefont {V.}~\bibnamefont {Makarov}},\ }\bibfield  {title} {\bibinfo {title} {Controlling single-photon detector id210 with bright light},\ }\href {https://doi.org/10.1364/OE.27.032253} {\bibfield  {journal} {\bibinfo  {journal} {Optics Express}\ }\textbf {\bibinfo {volume} {27}},\ \bibinfo {pages} {32253} (\bibinfo {year} {2019})}\BibitemShut {NoStop}%
\end{thebibliography}%

\end{document}